\documentclass[12pt,preprint]{aastex}

\usepackage{rotating}
\usepackage{pdflscape}
\newcommand{\msun}{M$_\odot$}

\newcommand{\joz}{\mbox{$J=1 \rightarrow 0$}}
\newcommand{\jto}{\mbox{$J=2 \rightarrow 1$}}

\newcommand{\omm}{\mbox{$\lambda=1.3$ mm}}
\newcommand{\tmm}{\mbox{$\lambda=2.7$ mm}}

\newcommand{\cojto}{$\textrm{CO}~J=2\rightarrow1$}

\newcommand{\wktextbf}{\textbf}
\renewcommand{\textbf}{\textrm}

\slugcomment{Draft revised in 2015 April}

\shorttitle{Resolving Protoplanetary Disks at Millimeter Wavelengths}
\shortauthors{Kwon et al.}

\begin{document}

\title{Resolving Protoplanetary Disks at Millimeter Wavelengths by CARMA}

\author{Woojin Kwon\altaffilmark{1,2,3} and Leslie W. Looney\altaffilmark{1}
and Lee G. Mundy\altaffilmark{4} and  William J. Welch\altaffilmark{5}}
\email{wkwon@kasi.re.kr}
\altaffiltext{1}{Astronomy Department, University of Illinois, 1002
West Green Street, Urbana, IL 61801}
\altaffiltext{2}{SRON Netherlands Institute for Space Research,
Landleven 12, 9747 AD Groningen, The Netherlands}
\altaffiltext{3}{Korea Astronomy and Space Science Institute,
776 Daedeok-daero, Yuseong-gu, Daejeon 305-348, Republic of Korea}
\altaffiltext{4}{Astronomy Department, University of Maryland,
College Park, MD 20742}
\altaffiltext{5}{Astronomy Department and Radio Astronomy Laboratory,
University of California, Berkeley, CA 94720}

\begin{abstract}
We present continuum observations at \omm\ and 2.7 mm using the
Combined Array for Research in Millimeter-wave Astronomy (CARMA)
toward six protoplanetary disks in the Taurus molecular cloud: CI Tau, DL
Tau, DO Tau, FT Tau, Haro 6-13, and HL Tau.  We constrain physical
properties of the disks with Bayesian inference using two disk models;
flared power-law disk model and flared accretion disk model.
Comparing the physical properties,
we find that the more extended disks are less flared and that the dust opacity
spectral index ($\beta$) is smaller in the less massive disks.
In addition,
disks with a steeper mid-plane density gradient have
a smaller $\beta$, which suggests that grains grow and radially move.
Furthermore, we compare the
two disk models quantitatively and find that
the accretion disk model provides a better fit overall.
We also discuss the possibilities of substructures on 
three extended protoplanetary disks.
\end{abstract}


\section{Introduction}
\label{sec_intro}
Circumstellar disks of young stellar objects (YSOs), particularly
at the stages of Class II and III (the so-called T Tauri stars),
are often called protoplanetary
disks, since they are expected to form planets.  These disk structures
have been studied by spectral energy distribution (SED) over the past
few decades \citep[e.g.,][]{beckwith1990, andrews2005}.  However,
in order to derive the detailed physical properties radio interferometry
at (sub)millimeter wavelengths is required, since the disks are
cold ($T \sim 30$ K) and small ($D \lesssim 1 \arcsec$).  To
date about two dozen protoplanetary disks have been studied by
interferometric observations with sub-arcsecond resolution
\citep[e.g,][]{2009ApJ...700.1502A, 2009ApJ...701..260I, 2011A&A...529A.105G}.
Recently the Atacama Large Millimeter/submillimeter
Array (ALMA) started to present excellent results on
transition disks, which are more evolved than protoplanetary disks,
with the unprecedented sensitivity \citep[e.g.,][]{2013Sci...340.1199V,
2014ApJ...787...42C}.
Also, detailed studies of grain growth in protoplanetary disks beyond
millimeter sizes
has been carried out with the expanded Karl G. Jansky Very Large Array (JVLA)
\citep[e.g.,][]{2012ApJ...760L..17P}.
\textbf{In addition, the long baseline science verification (SV) data of ALMA
have recently shown unprecedented details of substructures
toward HL Tau, which is one of our sample.}

In the previous modeling studies of protoplanetary disks using SED
or early interferometric data, a power-law disk model with sharp
inner and outer edges was utilized \citep[e.g.,][]{beckwith1990,
1996ApJ...464L.169M}.  Although the assumption is quite reasonable
since the minimal-mass solar nebula shows a power-law density
distribution \citep{1977Ap&SS..51..153W} and numerical simulations
of circumstellar disks also show a power-law density distribution
\citep[e.g.,][]{2009MNRAS.397..657A}, it has
no fundamental physics behind it.  In addition,
the model does not explain the difference between disk sizes
detected in dust continuum (compact) and CO spectral line (extended).

The viscous accretion disk model provides a more physically motivated
model, based on accretion with conservation of angular momentum
\citep[e.g.,][]{1981ARA&A..19..137P}.  The model has a density
distribution with a form of a power-law tapered by an exponential
function.
One complication in this model is that the sources of viscosity in
the disk are not well understood and may vary with radius and disk
mass.
\citet{2008ApJ...678.1119H} argued that the
viscous accretion disk model explains both dust continuum and gas
spectral line data better than the power-law disk model by qualitative
comparison.
In this study, we employ both flared power-law disk and
flared viscous accretion disk models to  investigate disk properties;
we compare the two disk models in order to determine which model
better fits our millimeter data.   Although some studies have modeled
the dust grain properties with radius in the disk using submillimeter
to centimeter data \citep[e.g.,][]{2012ApJ...760L..17P}, we assume
constant dust properties in the disk due to our relatively limited
wavelength coverage.

Bayesian inference is used to determine the model parameter distributions
that fit the observational data, instead of the commonly used
$\chi^2$ method.  The Bayesian approach provides a basis of probability
theory to compare two models as well as to achieve parameter
probability distributions, which are not given by the $\chi^2$
method.  Although the $\chi^2$ method has been used widely, a couple
of circumstellar disk studies have been carried out utilizing
Bayesian inference \citep{lay1997,2009ApJ...701..260I,2009PhDT........18K}.
No model comparison has been attempted for
circumstellar disks so far in the Bayesian approach.
\citet{2010ApJ...714.1746I} obtained slightly smaller reduced
$\chi^2$ values toward their \omm\ data of DG Tau and RY Tau for
the accretion disk model than the power-law disk model.  On the
other hand, \citet{2011A&A...529A.105G} and \citet{2002ApJ...581..357K}
also utilized both models.  However, their studies used thin disk
models and a simple power-law temperature
distribution.  Our disk models are flared and our temperature
distributions are obtained by a Monte Carlo radiative transfer code,
which are more realistic.

In this paper, we present observational and modeling
results of our protoplanetary
disk survey, which has taken high angular resolution data
at \omm\ and 2.7 mm using the Combined Array for Research in
Millimeter-wave Astronomy (CARMA) providing high image fidelity.
First, brief introductions
of the targets are given in Section \ref{sec_targets}\ and CARMA
observations and data reduction are discussed in the following
section.  Afterward the two disk models of flared power-law disks
and viscous accretion disks are explained,
followed by observational and fitting results and discussion in Section
\ref{sec_results}.  Finally, we summarize the results.

\section{Targets}
\label{sec_targets}

The circumstellar disk targets of this study
(CI Tau, DL Tau, DO Tau, FT Tau, Haro 6-13, and HL Tau;
Table \ref{tab_targets})
are located in the Taurus
molecular cloud, a well-known, nearby
star forming region.  The distance has been determined by various
methods \citep{2004AJ....127.1029R}: e.g., star counting
\citep[e.g.,][]{1938ApJ....88..209M}, optical extinction \citep[e.g.,]
[$140\pm 10$ pc]{1994AJ....108.1872K}, parallax measurement \citep[e.g.,
Hipparcos astrometric data:][]{1999A&A...352..574B}, and protostellar
rotational properties \citep[][$152\pm 10$ pc]{1997A&A...322..825P}.
The measured distances to the Taurus molecular cloud are somewhat
different in methods and in regions. For example,
\citet{1999A&A...352..574B} reported three regions of $125^{+21}_{-16}$
pc, $140^{+16}_{-13}$ pc, and $168^{+42}_{-28}$ pc, using Hipparcos
astrometric data.  However, 140 pc is within uncertainties of most
of the ranges, so it is adopted for all 6 targets in this study.
Note that the distance affects the physical sizes and mass estimates.

Our targets
were chosen from the sample of \citet{beckwith1990},
which observed 86 YSOs in the Taurus molecular cloud at \omm\ using
the IRAM 30 m telescope.
They were also included in the submillimeter survey at
$\lambda= 450$ and $850~\mu$m carried out by \citet{andrews2005}
using SCUBA on the James Clerk Maxwell Telescope.
On the other hand, \citet{1995ApJS..101..117K} compiled
published IR and optical observations with supplementary IR
observations in the cloud.
In addition, \citet{2009ApJ...703.1964F} reported
$\lambda \approx 10~\mu$m feature strengths (equivalent width of
the $10~\mu$m feature, EW($10~\mu$m)) and spectral index ($n_{13-31}$)
between $\lambda=13$ and $31~\mu$m of T Tauri disks in the Taurus
molecular cloud including our targets except HL Tau, as well as two
other star forming regions using the {\it Spitzer} Space Telescope
IRS data.  They utilized these
values to indicate dust evolution; for example, smaller $n_{13-31}$
values represent disks with more dust settlement (less flaring) and
EW($10~\mu$m) shows the amount of small dust grains ($< 5~\mu$m),
which means that smaller EW($10~\mu$m) indicates more evolved dust.
Selected information from the above studies is summarized in Table
\ref{tab_targets}.  More detailed studies have been carried out
toward these targets and some selected examples are introduced in
the following.

{\bf \underline{CI Tau}}
was observed by \citet{beckwith1990} with a continuum flux of $190\pm17$
mJy at \omm.  In addition, variability in optical spectra has been
detected, which can be interpreted as a temporary obscuration of a
local hot region \citep{1999MNRAS.304..367S}.  \citet{andrews2007}
observed CI Tau using the Submillimeter Array (SMA)
at $\lambda=880~\mu$m with a moderate
angular resolution ($> 1 ''$).  By model fitting of SED data and
visibility data averaged in annulus, they constrained physical
properties such as density and temperature distributions, size, and
mass. \textbf{\citet{2011A&A...529A.105G} also studied this target by about $0.5''\times0.3''$
angular resolution data at two wavelengths of the Plateau de Bure Interferometer 
(PdBI) and two disk models: power-law and accretion disk models.}

{\bf \underline{DL Tau}}
has been detected with a continuum flux of $230\pm14$ mJy by IRAM observations
\citep{beckwith1990}.  It has also been detected in the \cojto\ line, and
the velocity field was shown \citep[e.g.,][]{1995AJ....109.2138K}.
\citet{2000ApJ...545.1034S} studied the kinematics using PdBI data
in \cojto.  They showed that the disk is in Keplerian rotation and
estimated the protostellar mass as $0.72\pm0.11$ \msun\ assuming the
inclination determined by the continuum image ($49\pm3 \degr$).
\citet{2002ApJ...581..357K} has studied the disk structure by fitting
SEDs and Nobeyama Millimeter Array data taken at $\lambda = 2$ mm
with $1''$ resolution.  They employed a viscous accretion disk as
well as a power-law disk model.  \citet{andrews2007} also
studied the DL Tau disk structure using SMA data at \omm\ with about
$2''$ resolution. \textbf{This target was also studied by \citet{2011A&A...529A.105G}
using PdBI.}

{\bf \underline{DO Tau}}
has a continuum flux of $136\pm11$ mJy in the \citet{beckwith1990} observations.
\citet{1995AJ....109.2138K} have reported \cojto\ observational
results .  The line profile has a much broader feature in the
blueshifted region than in the redshifted region, which indicates
that it needs infall and bipolar outflow components.  DO Tau is also one
of the sample in the studies of \citet{2002ApJ...581..357K}.

{\bf \underline{FT Tau}} has a continuum
flux measured in \citet{beckwith1990} of $130\pm14$ mJy at \omm.  In previous
survey observations at optical wavelengths, however, the spectral
type of FT Tau was not determined probably due to an undetermined
luminosity, although the surveys included the target.  Therefore,
there is no good estimate of protostellar mass and age, which are
normally determined by its effective temperature and luminosity.
\citet{2007ApJS..169..328R} reported wide ranges of stellar parameters
by SED fitting: for example, protostellar temperature between 3060
and 5013 K and mass between 0.11 and 2.03 \msun.
\citet{andrews2007} observed FT Tau with an angular resolution
of $1.7'' \times 1.1''$ at $\lambda=880~\mu$m using SMA.  Although
they did not resolve the disk structure, they constrained disk
properties by model fitting with the data and SEDs.
\textbf{The study of \citet{2011A&A...529A.105G} using PdBI included
this target taken at about $0.5''\times0.3''$ angular resolution.}

{\bf \underline{Haro 6-13}}
has been detected with a continuum flux of $124\pm13$ mJy by
\citet{beckwith1990}.  \citet{2009ApJ...701..698S} observed the
target in CO \jto\ and \joz\ as well as \omm\ and 2.7 mm continuum
using PdBI.  They estimated the protostellar mass as $1.0\pm0.15$
\msun, based on the rotational velocity field.  They discussed that
the difference from the mass estimated by the evolutionary track
on the HR diagram ($\sim0.6$ \msun) may be due to limited number
of channel maps for the CO data model fitting.
\textbf{The PdBI data of \citet{2009ApJ...701..698S} were also used in 
\citet{2011A&A...529A.105G}.}

{\bf \underline{HL Tau}}
has been studied very broadly over the last two decades, as it is
a bright T Tauri star.  It has been studied by submillimeter and
millimeter imaging of interferometers \citep[e.g.,][and references
therein]{2011ApJ...741....3K}, as well as molecular line observations.
In particular, by a modeling of the viscous accretion disk model
to millimeter CARMA data and SED data, we found that HL Tau experiences
stratified grain settlement and has an outer disk region gravitationally
unstable \citep{2011ApJ...741....3K}.  In fact, this paper uses the
same data set of HL Tau but we employ two disk models and a better
temperature distribution. \textbf{HL Tau has also been studied by
PdBI \citep{2011A&A...529A.105G} and has recently been
observed by ALMA at three wavelengths (ALMA Band 3, 6, and 7) with
an excellent angular resolution down to $\sim0.02''$.}

\section{Observations and Data}
\label{sec_obs}

We carried out a survey of protoplanetary disks using CARMA,
which is a heterogeneous array consisting of six 10.1-m, nine 6.4-m,
and eight 3.5-m antennas.  For this project the 3.5-m antennas were
not used for imaging.
All targets were observed in A and/or B configurations
(see Table \ref{tab_diskobs}) at \omm\ and/or 3 mm, which provide
high angular resolution better than $0.3$\arcsec\ at \omm.  Table
\ref{tab_diskdata} provides details on the final combined data,
their synthesized beam sizes and noise levels.

In addition to the most extended A and B configurations,
we have observed the protoplanetary disks in C and D configurations,
in order to detect large scale structures as well
as small scales.  The highest angular resolution (the smallest
synthesized beam size) is $\sim$0.13\arcsec\ at \omm\
toward HL Tau.  The angular resolutions for the other targets are
a bit less ($\sim$0.3\arcsec).  The {\it uv} coverage of our data
are summarized in Table \ref{tab_diskdata}.  As shown in the table,
the minimum {\it uv} distance is at most 17 kilo-wavelengths,
which means that our large scale sensitivity is at least 10 arcseconds.
As the target disks are less than a few arcseconds
in size, there is no resolving-out flux issue in the study, from
which interferometric observations typically suffer for extended
targets.

The largely extended arrays (A and B configurations) require
stable atmospheric conditions due to long baselines.  It is
challenging to obtain such good weather conditions even at the CARMA
site.  To verify our data quality, therefore, we have observed an
additional calibrator (a test calibrator) with a gain calibrator.
Since a test calibrator is chosen to be a point source nearby our
targets, data quality after applying gain solutions can be examined
by the test calibrator's shape and flux.  All of our A and B
configuration data and some of the compact array data have such a
test calibrator.  Most of our data have built a point or point-like
image of a test calibrator, which means successful calibration.
Here ``point-like'' means that the deconvoluted size of a test
calibrator is less than half of a synthesized beam in both major
and minor axes.  A few of the worst data sets have a test calibrator
size comparable to the synthesized beam and those sets are indicated
in Table \ref{tab_diskobs}.  In addition, we examined the flatness
of the {\it uv} amplitude slopes of the gain calibrators.  Since
{\it uv} data are the Fourier transform of an image, a gain calibrator
of a point source should have a flat amplitude versus {\it uv}
distance.  If it is not achieved, i.e., amplitudes at long {\it uv}
distance regions (small scales, central regions) drop, the target
data are likely to be biased, resulting in an apparent shallower
gradient of density and/or temperature.  All of our data sets have
a flat amplitude slope of gain calibrators after calibration, except
a data set taken in B configuration at \omm\ toward DO Tau, which
has about 25\% variation in the middle.  However, it does not seem
to affect the results significantly, as the variation is acceptable
in magnitude and is positioned in the middle of {\it  uv} coverage.
Therefore, we include that data as well with caution, in order to
have long baseline data.

CARMA itself has a special technique for calibration of atmospheric
perturbation, called CARMA Paired Antenna Calibration System (CPACS)
\citep{PACS}.  In large antenna spacings such as A and B configurations,
the 3.5 m antennas are paired with boundary 10.4 m or 6.1 m antennas,
which are elements of long baselines.  The 3.5 m antennas are always
observing a calibrator at 30 GHz, while the other antennas are
observing science targets and calibrators.  During the calibration
of data, the phase of the 3.5 m antenna gain solution is used to
correct the phase delay (phase difference) caused by short atmospheric
perturbations ($\sim$4--20 seconds) on the paired antennas.  Since
visibility phase depends on time delay and wavelengths/frequencies,
the phase of the gain solution is scaled by the ratio of observation
frequencies to 30 GHz.  The A configuration data of HL Tau at \omm\
and the A configuration data of FT Tau at \tmm\ are calibrated using
CPACS.  The calibrators used for CPACS are separated from HL Tau and
FT Tau by $9 \degr$ and $13 \degr$, respectively, and the improvement
of calibrated data is about 10--20\% in terms of image noise levels
and the size, flux, and peak intensity of test calibrators.

We obtained two wavelength data (\omm\ and
2.7 mm) in order to better constrain the dust properties.
The properties constrained by multi-frequency
data mainly depend on absolute flux calibration.  To minimize the
bias induced by flux calibration uncertainty, a good flux calibrator
(e.g., Uranus) was used (Table \ref{tab_diskobs}).  In addition,
gain calibrator fluxes of all tracks have been compared with each
other in the time basis.  Rapid variation of gain calibrator fluxes
(e.g., 50\% increase or decrease within a few days) is unrealistic,
although quasar gain calibrators are intrinsically variable.  For
comparison, the CARMA flux catalog and the SMA calibrator list have
been taken into account, as well as the gain calibrator values
bootstrapped from flux calibrators.  In addition,
different array-configuration data of the same targets
have been compared at common {\it uv} distances.  Sub/millimeter
dust emission from T Tauri disks is not variable over a period of
a few years, so amplitudes at common {\it uv} positions should be
comparable even in different configuration arrays.
Actually, this comparison is a crucial step when combining various
array-configuration data.  As a result, we presume absolute flux
calibration uncertainty of our T Tauri disk data is less than the
uncertainty assumed in other studies, which use only one or two
tracks in an array-configuration for each frequency data:
10\% at \omm\ and 8\% at \tmm\ \citep[e.g.,][]{2011ApJ...741....3K}.

We also considered proper motions of our targets when combining our
data, as they have been taken over a few year period.  However,
since proper motions of all targets have not yet been well estimated,
we decided not to correct them blindly.  Instead, we estimated the
proper motion based on the data themselves. In the case of Haro
6-13, whose peaks show an offset in the data set, we matched the
peak positions.  As a result, we shifted the 3 mm data sets by
(R.A., Dec.) = ($0.0105''$, $0.055''$).  In addition, for HL Tau,
whose data have the highest angular resolution and whose proper
motion is well measured, we compensated for the
proper motion \citep{2011ApJ...741....3K}.

MIRIAD \citep{sault1995} has been employed to calibrate and map
data.  Individual tracks have been calibrated and verified separately
then combined in the invert-Fourier transform step to make maps.
The sensitivity and emphasized scales in the maps depend on the
visibility data weighting scheme.  Natural weighting gives the
highest signal-to-noise and the worst angular resolution, since
visibility data points are sparse at large {\it uv} distances and
visibilities at larger {\it uv} distances are noisier due to
atmospheric turbulence.  On the other hand, uniform weighting
emphasizes small scales best although the signal-to-noise is worst.
For intermediate weighting, Briggs introduced a robust parameter
\citep{briggs1995}: 2 of the parameter for weighting close to the
natural weighting and -2 close to the uniform weighting.  In order
to have small structures reasonably emphasized, the weighting scheme
with a robust value of 0 has been applied.

\section{MODELING}
\label{sec_diskmodel}
We employ two disk models to estimate disk physical parameters:
a viscous accretion disk model \citep[e.g.,][]{1981ARA&A..19..137P}
and a power-law disk model \citep[e.g.,][]{beckwith1990}.
The outer boundary of the two disk models are described differently.
While a sharp outer radius is associated with the power-law disk
model, a characteristic radius is introduced in the viscous accretion
disk model, which has a power-law density structure tapered by an
exponential function.  We also consider vertical structures for
investigating disk thickness and flare degree.  For our models, we
use cylindrical coordinates, and the disk models are axisymmetric.
Therefore, the physical properties are expressed in $R$ and $z$
coordinates: e.g., density structures $\rho(R,z)$.  On the other
hand, $r$ indicates a direct distance from the disk center, i.e.,
$r^2 = R^2 + z^2$. \textbf{The model emission is calculated with a logarithmic
radial grid to ensure accurate values in the inner region of the disk.}

\subsection{Power-law Disk Model}
\label{sec_pwmodel}
The power-law disk model has a power-law density distribution with
a sharp edge\footnote{\textbf{We use $s$ for a volume density distribution
and $p$ for a surface density in order to minimize confusion with other studies.
Note that this notation is different from \citet{2011ApJ...741....3K}.}},
\begin{eqnarray}
\rho(R,z) &=& \rho(R,0)~\textrm{exp}[{-(z/H(R))^2}] \nonumber \\
   &=& \rho_0 \Big(\frac{R}{R_0}\Big)^{-s} \textrm{exp}[{-(z/H(R))^2}],
\end{eqnarray}
where $H(R)$ is a scale height along radius set by a power-law function:
\begin{equation}
\label{eq_H}
H(R) = b_{t} H_0 \Big(\frac{R}{R_0}\Big)^h.
\end{equation}
Here $H_0 = (2kR_0^3T_0/GM_*\bar{m})^{0.5}$ is the
hydrostatic equilibrium case between the vertical component of the
protostellar gravity and the local pressure of $T(R_0,0)=T_0$.
We ignore self gravity of the disk mass and assume that dust
temperature and gas kinetic temperature are the same.
In the $H_0$ expression, $k$ is the Boltzmann's constant, G
is the gravitational constant, $M_*$ is the central protostellar
mass, and $\bar{m}$ is the mean molecular mass: $\bar{m} = \mu m_H
= 2.4 m_H$.
Note that disks in the hydrostatic equilibrium and in the
optically thin condition have $b_t = 1$ and $h = 3/2 - q/2 \approx 1.25$,
where $T(R,0) \propto R^{-q}$.

As the surface density is obtained by the integration of the density
distribution in $z$,
the surface density is also a power law with a power index
of $p=s-h$:
\begin{eqnarray}
\label{eq_surfacedensity}
\Sigma(R) &=& \int_{-\infty}^{\infty} \rho(R,z) dz \nonumber \\
       &=& \rho(R,0) \sqrt{\pi} H(R) \nonumber \\
       &=& \rho_0 \sqrt{\pi} b_t H_0 \Big(\frac{R}{R_0}\Big)^{-p}.
\end{eqnarray}
In this model we also have a sharp inner radius ($R_{in}$) in addition to
the outer radius ($R_{out}$).
Therefore, $\rho_0$ is expressed
with a disk total mass ($M_{disk}$),
\begin{eqnarray}
M_{disk} &=& \int^{R_{out}}_{R_{in}} \int^{\infty}_{-\infty}
\rho(R,z)dz ~2\pi R~dR \nonumber \\
 &=& \rho_0 \frac{2\pi^{1.5} R_0^2 b_t H_0}{2-p} \Big[
\Big(\frac{R_{out}}{R_0}\Big)^{2-p} - \Big(\frac{R_{in}}{R_0}\Big)^{2-p}
\Big] \nonumber \\
\rho_0 &=& M_{disk} \frac{2-p}{2\pi^{1.5} R_0^2 b_t H_0} \Big[
\Big(\frac{R_{out}}{R_0}\Big)^{2-p} - \Big(\frac{R_{in}}{R_0}\Big)^{2-p}
\Big]^{-1}
\end{eqnarray}

Although we consider a power-law temperature distribution for the
disk scale height, we use a temperature distribution calculated in
a given disk structure by the RADMC 3D code \citep{dullemondcode}
for more realistic modeling with a temperature gradient along the
vertical direction.  However, we did not update the scale height
based on the calculated temperature distribution.  This decoupling
of the disk structure from the temperature distribution is justified,
because dust grains settle to the mid-plane rather than
mix with gas as they grow \citep[e.g.,][]{2009ApJ...700.1502A}.

The RADMC 3D code provides two dimensional temperature distributions,
but it requests a large number of photons for a smoothly varying
distribution particularly in the case of thin disks with a high
optical depth.  Therefore, instead of simply using the temperature
distributions out of the RADMC 3D code, we make a functional form
weighting two functions that fit to the mid-plane ($T_m$) and surface
temperature distributions ($T_s$),
\begin{equation}
T(R,z) = W T_m(R,0) + (1 - W) T_s(r).
\end{equation}
Based on qualitative comparisons, we chose the weighting function
$W=\textrm{exp}[-(z/H(R))^2]$.

The dust opacity spectral index $\beta$ is one of the most interesting
parameters in this study, which represents dust grain sizes
\citep{Draine2006}: $\kappa_\nu = \kappa_0 (\nu/\nu_0)^\beta$.
Indeed, in order to constrain $\beta$, we fit the two wavelength data
simultaneously.  Therefore, we assume $\kappa_0 = 0.01$ \mbox{cm$^2$
g$^{-1}$} at 230 GHz and $\beta$ is determined between \omm\ and
2.7 mm.  The $\kappa_0$ is based on the case of ice mantle grains
following a MRN size distribution \citep{1977ApJ...217..425M}\ in
\citet{ossenkopf1994}.  In addition, we assume a gas-to-dust mass
ratio of 100.  The assumed $\kappa_0$ corresponds to the case of
$\kappa_\nu = 0.1 (\nu/1200~GHz)^\beta$ with $\beta=1.39$
\citep[e.g.,][]{hildebrand1983,beckwith1991}.  The uncertainty of
$\kappa_0$ is large, around factor of two \citep{ossenkopf1994}.
For the opacity as an input of RADMC when determining disk temperature
distributions we simply consider the combination of an opacity curve
at infrared wavelengths in the case of silicate-carbon grains in
the MRN size distribution and a power-law at millimeter wavelengths.

\subsection{Viscous Accretion Disk Model}
In addition to the power-law disk model,
we utilize the viscous accretion disk model for fitting our data.
The viscous accretion disk model has a density distribution
of a power-law tapered by an exponential function
\citep[e.g.,][]{1981ARA&A..19..137P},
\begin{eqnarray}
\label{eq_accsurfacedensity}
\rho(R,z) &=& \rho_0 \Big(\frac{R}{R_c}\Big)^{-s}
\textrm{exp}\Big[\Big(\frac{R}{R_c}\Big)^{2-s+h}\Big]~
\textrm{exp}\Big[-\Big(\frac{z}{H(R)}\Big)^2\Big].
\end{eqnarray}
The viscous accretion disk model is expressed as
the surface density distribution \citep[e.g.,][]{2009ApJ...700.1502A},
\begin{eqnarray}
\Sigma(R) &=& \int_{-\infty}^{\infty} \rho(R,z) dz \nonumber \nonumber \\
&=& \rho_0 \sqrt{\pi} b_t H_0 \Big(\frac{R_c}{R_0}\Big)^h
\Big(\frac{R}{R_c}\Big)^{-s+h}
\textrm{exp}\Big[-\Big(\frac{R}{R_c}\Big)^{2-s+h}\Big] \nonumber \\
&=& \Sigma_0 \Big(\frac{R}{R_c}\Big)^{-\gamma}
\textrm{exp}\Big[-\Big(\frac{R}{R_c}\Big)^{2-\gamma}\Big],
\end{eqnarray}
where $R_c$ is a characteristic radius.
Note that $\Sigma_0 = \rho_0 \sqrt{\pi} H(R_c)$ and $\gamma = s-h$.
Like the power-law disk model, we decouple the scale height distribution
from the temperature distribution, and the density distribution is
expressed with the total disk mass,
\begin{eqnarray}
M_{disk} &=& \int_{R_{in}}^{\infty} \int_{-\infty}^{\infty}
\rho(R,z) dz ~2\pi R~dR \nonumber \\
 &=& \rho_0 \frac{2\pi^{1.5} R_c^2 H(R_c)}{2-\gamma}
\textrm{exp}\Big[-\Big(\frac{R_{in}}{R_c}\Big)^{2-\gamma}\Big] \nonumber \\
\rho_0 &=& M_{disk} \frac{2-\gamma}{2\pi^{1.5} R_c^2 H(R_c)}
\textrm{exp}\Big[\Big(\frac{R_{in}}{R_c}\Big)^{2-\gamma}\Big].
\end{eqnarray}
Everything except the density distribution is set in the same way
to the power-law disk model. Note that the characteristic
radius ($R_c$) is introduced instead of the outer radius ($R_{out}$).

\subsection{Model Parameters}
\label{sec_param}
There are 7 free parameters for each disk model:
$s$, $\beta$, $M_{disk}$, $R_{in}$, $R_{out}$ (or $R_c$),
$\theta_i$, and $PA$. \textbf{In addition, there are two parameters
related to the disk scale height, $h$ and $b_t$, which are weakly constrained
by the observations and will be handled as ``controlled parameters''
in the fitting.}
The $\theta_i$ is an
inclination angle of a disk (a face-on disk with $\theta_i=0\degr$)
and $PA$ is a position angle of a disk measured eastward from the
north.
The $R_{out}$ and $R_c$ are for the cases
of power-law disk and viscous accretion disk models,
respectively.
For the viscous accretion disk model, an outer disk radius is not
employed.  However, the modeling disks are cut off at 5 $R_c$
due to limitation in image size and
for simplicity in integration.
Some modeling studies of circumstellar disks have presumed sublimation
radii of dust grains \citep[e.g.,][]{2009ApJ...700.1502A}, but it
is possible that inner radii ($R_{in}$) are determined by other
effects such as an unresolved binary companion, formed planets,
etc.  Therefore, we leave the inner radii as a free parameter, \textbf{but fitted values 
can be model-dependent and are likely unreliable due to systematic
uncertainties in the observation, primarily the resolution.}

The sign of inclinations are determined by
our spectral line data in CO, which were included in our observations.
We only detected CO in data from the C and D configurations,
which traces the bipolar outflows as well as a disk structure.
This allowed us to estimate the disk inclination, at least the sign.
CI Tau and FT Tau, which we have not detected any CO, are set to
have a plus sign of inclination.
All the inclination signs are consistent with previous
studies \citep[e.g.,][]{2011A&A...529A.105G}.

The two controlled parameters, $b_t$ and $h$, are used to
investigate the disk vertical structure.
As shown in equation \ref{eq_H}, $b_t$ and $h$ are disk scale heights
with respect to the hydrostatic equilibrium cases and disk flare
indexes.  In order to figure
out whether preferred disks are thin or thick and more or less flared,
we carry out 25 separate models for each target: $b_t = \{0.6, 0.8,
1.0, 1.2, 1.4\}$ and $h = \{0.95, 1.10, 1.25, 1.40, 1.55\}$. 
\textbf{We adopted the 25 cases instead of having them as two
additional free parameters for maximizing the efficiency of our
modeling without losing our study goals.}

\subsection{Model Fitting Procedures}
\label{sec_fitting_proc}

First, a temperature distribution is obtained by the
RADMC 3D code in a given disk structure and fitted
as described in Section \ref{sec_pwmodel}.
Then, a disk model is built in two dimensional logarithmic grids
by solving the radiative transfer equation numerically along the
line of sight.  Note that the disk inclination is applied when
constructing the disk model, so the disk model must be two dimensional
and the radiative transfer equation is numerically integrated along
the third axis, line of sight.  No optically thin nor Rayleigh-Jeans
approximations are assumed.  Afterward, disk images are made by
interpolations of the disk models in two dimensional linear pixels.
The center positions and position angles (PAs) of the disks are
applied when making the disk images.

The disk images are multiplied by three different normalized primary
beams, which correspond to three different types of baselines in
CARMA: 10.4 m--10.4 m antennas, 10.4 m--6.1 m antennas, and 6.1
m--6.1 m antennas.  These three primary beam corrected images are
Fourier-transformed into model visibility maps.  Then, the actual
model visibilities are sampled in the {\it uv} coverage of observational
data by bi-linear interpolation of the model visibility maps.  Image
pixel sizes and image sizes are selected to obtain reasonable pixel
sizes and {\it uv} coverage of visibility maps.  For example, an
image pixel size of 9 AU and an image size of 1024 pixel by 1024
pixel are used, which provide {\it uv} coverage up
to 1600 $k\lambda$ ({\it uv} range: -1604 to +1604 $k\lambda$)
and a visibility map pixel of about 3.1 $k\lambda$, based on the
Nyquist theorem: for example, the {\it uv} range $\Delta_{uv} = 1/2
\delta_{image}$, where $\delta_{image}$ is the image pixel size.

Model fitting is done by comparing observational data with the model
data sampled along the observational {\it uv} coverage in
Bayesian inference \citep[e.g.,][]{gilks1996,mackay2003}.
Bayesian inference allows us to obtain the probability distribution
of disk properties \textbf{($m$) with given data $D$ and a given disk model $H$}: 
$P(m\mid D, H)$,
\begin{equation}
\label{eq_bayes}
P(m \mid D, H) = \frac{P(D \mid m, H) P(m \mid H)}{P(D \mid H)}.
\end{equation}
Here $P(m \mid D,H)$ is called the posterior probability,
$P(D\mid m,H)$ the likelihood, $P(m\mid H)$ the prior, and
$P(D\mid H)$ the evidence.  The evidence is also called
the marginal likelihood, which is used to compare multiple models,
since $P(D\mid H)=\int dm P(D\mid m,H)P(m\mid H)
\approx \Sigma~P(D\mid m,H)P(m\mid H)$.
Gaussian functions are employed for the likelihood,
since noise of interferometric data is a normal distribution,
$\prod_i \textrm{exp}[-(D_i-M_i)^2/2\sigma_i^2]$.
Theoretical uncertainties of interferometric visibility points are
typically estimated as
$\sigma = ({2k}/{A\eta})\sqrt{{T_1 T_2}/{2t_{int}\Delta\nu}}$
\citep[e.g.,][]{ISRA2001},
where k is Boltzmann's constant, A is antenna area, $\eta$ is the
efficiency of the antenna surface and correlator quantization, $T_1$
and $T_2$ are system temperatures of antennas corresponding to the
baseline, $\Delta\nu$ is bandwidth, and $t_{int}$ is integration
time.
However, the uncertainties do not represent decorrelation of
atmospheric turbulence, which depends on $\it{uv}$ distance.
Therefore, instead we use the standard deviation of the imaginary
components of self-calibrated gain calibrator data,
which are generally a few times larger than the theoretical estimate
\citep{2011ApJ...741....3K}.
Individual
wide bandwidth windows ($\Delta\nu \sim 500$ MHz)
are averaged and considered a visibility point.  The central channel
frequency of the window is used for computing its {\it uv} distance
in units of wavelength.  As the data are real numbers in image
space, i.e., the {\it uv} coverage is symmetric at the phase center,
the symmetric data points are added to the sample.  In addition,
the real and imaginary components of complex visibility data are
regarded as independent data.

Searching distributions of model parameters by the Metropolis Hastings
method has two parts.
The first part is finding a convergence region of parameters.
For a better performance, we start with 8 initial conditions,
which are randomly generated.
After running for a while, the initial conditions converge
into a parameter set
(normally two sets with different signs of inclination angles).
Annealing is also used to accelerate the convergence.
We start a fresh run from the convergent parameter set of the
expected inclination sign (refer to Section \ref{sec_param}) with
a smaller standard deviation of the proposal Gaussian functions,
for each combination of $b_t$ and $h$, 25 in total for each target.
The smaller standard deviation of the proposal functions
are set to have an acceptance rate of about 10--50 \%.
The GASDEV subroutine of the Numerical Recipes using the Box-Muller
method is used to produce random numbers in the Gaussian distribution
\citep{NRF77}.

\section{Results and Discussion}
\label{sec_results}
\subsection{Combined Data and Images}

\begin{figure}
%\begin{center}
\includegraphics[angle=0,scale=0.6]{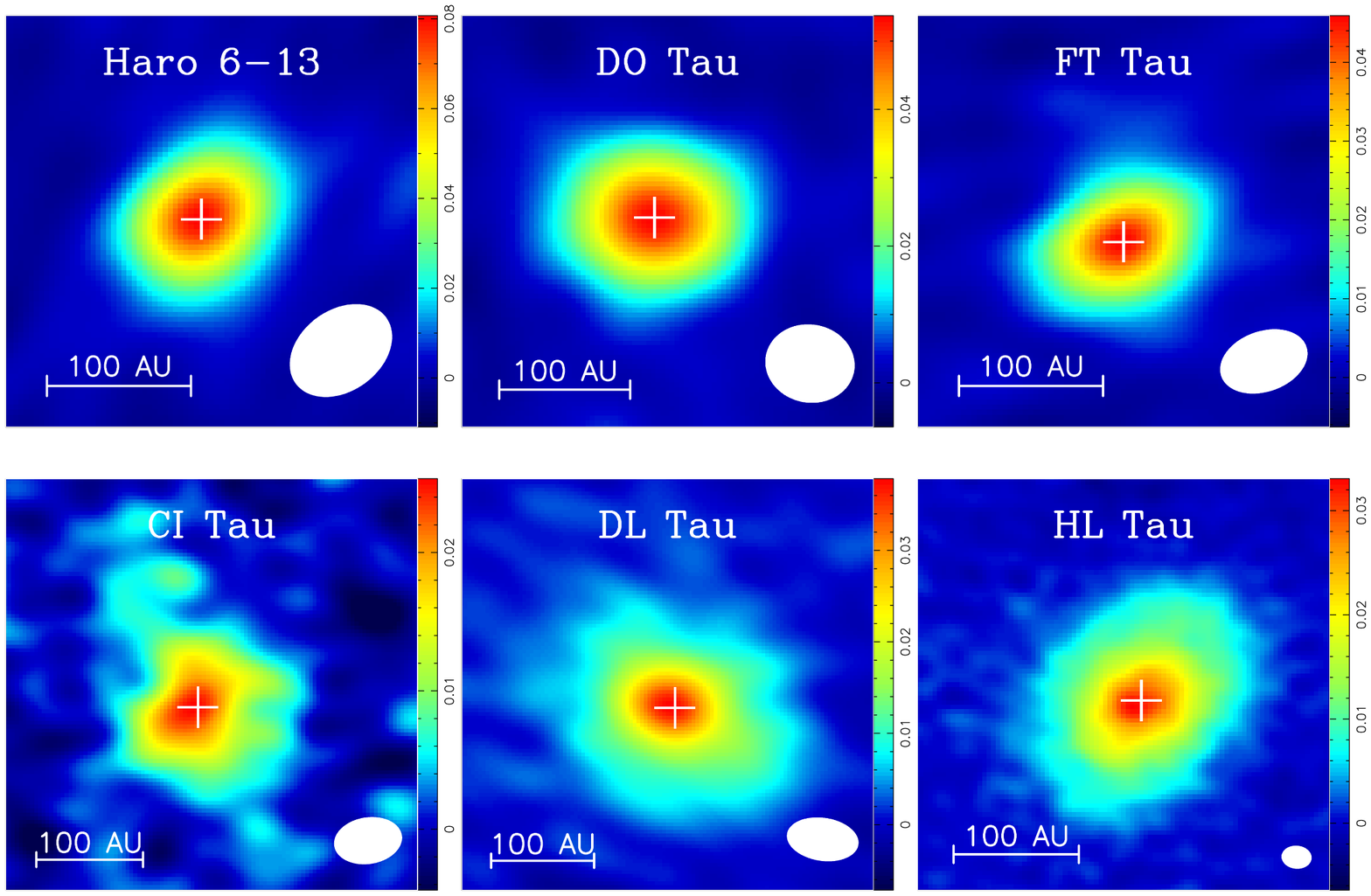}
\vspace{-2cm}
\caption{Protoplanetary disk images in $\lambda=1.3$ mm continuum.
The color scales are in units of \mbox{Jy beam$^{-1}$}, and the synthesized
beams and the 100 AU bars are marked at the bottom of each panel.
\label{fig_diskimages}}
%\end{center}
\end{figure}

Figure \ref{fig_diskimages} shows $\lambda=1.3$ mm continuum images
of the six protoplanetary disks; a physical scale bar of 100 pc and
the data angular resolution (synthesized beam) are marked.
Note that the three disks in the bottom panels
(CI Tau, DL Tau, and HL Tau) are extended, while the three disks
in the upper panels (Haro 6-13, DO Tau, and FT Tau) are relatively
compact. In addition, CI Tau has a weak non-axisymmetric feature
north-north-east to south-south-west.

The continuum noise levels that we achieved after combining all
configuration data are listed in Table \ref{tab_diskdata} with disk
total fluxes.  In addition, Table \ref{tab_diskdata} has disk center
positions estimated by a Gaussian fit (the IMFIT task of MIRIAD).
Table \ref{tab_targets} has phase center coordinates,
so one can simply compute the actual disk centers: e.g., HL Tau,
R.A. (J2000) = 04:31:38.418, Dec. (J2000) = +18:13:57.37.

\begin{figure}
\begin{center}
\begin{tabular}{c@{\hspace{-0.7cm}}c}
\vspace{-1.0cm}
\includegraphics[angle=270,scale=0.3]{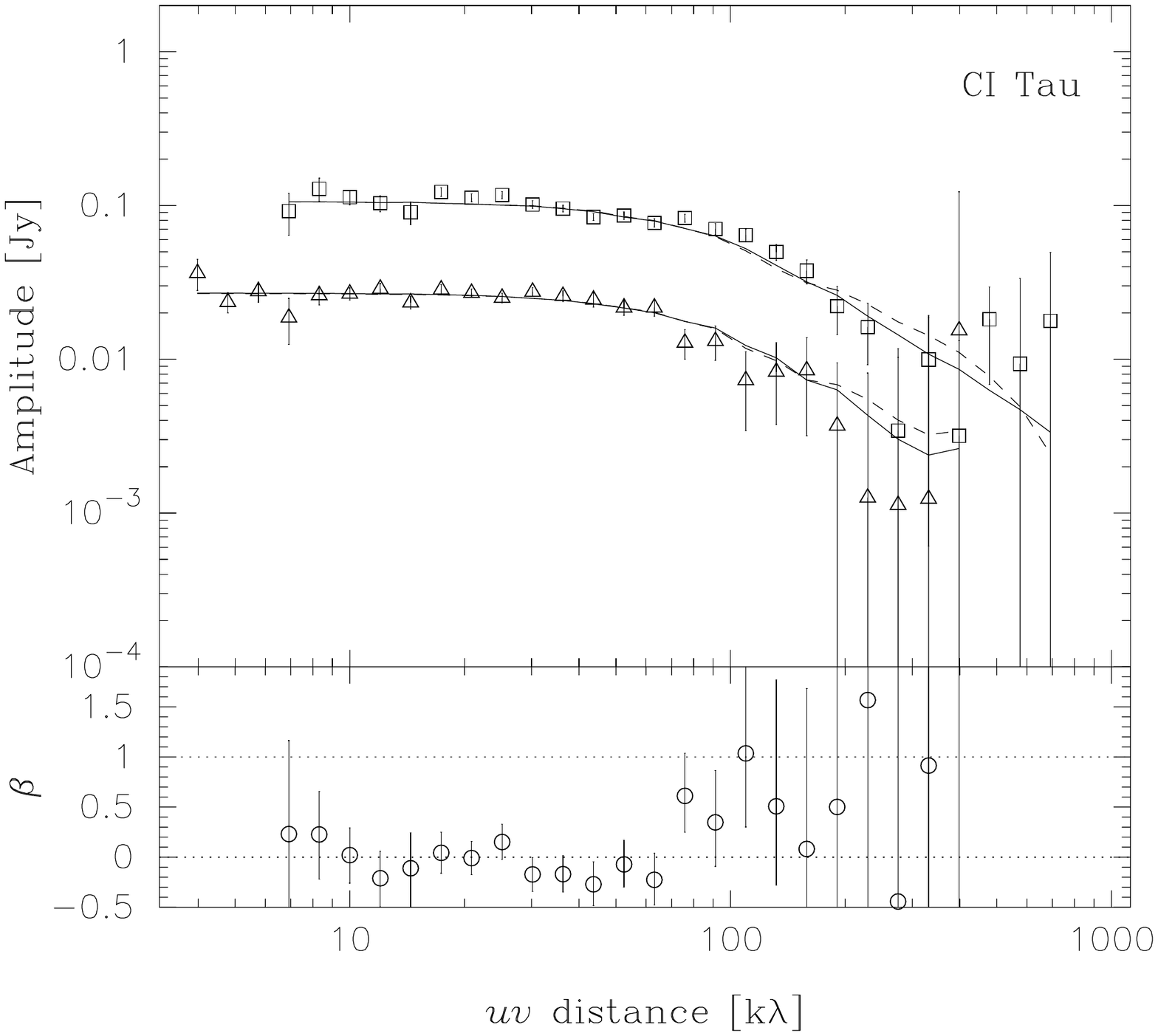} &
\includegraphics[angle=270,scale=0.3]{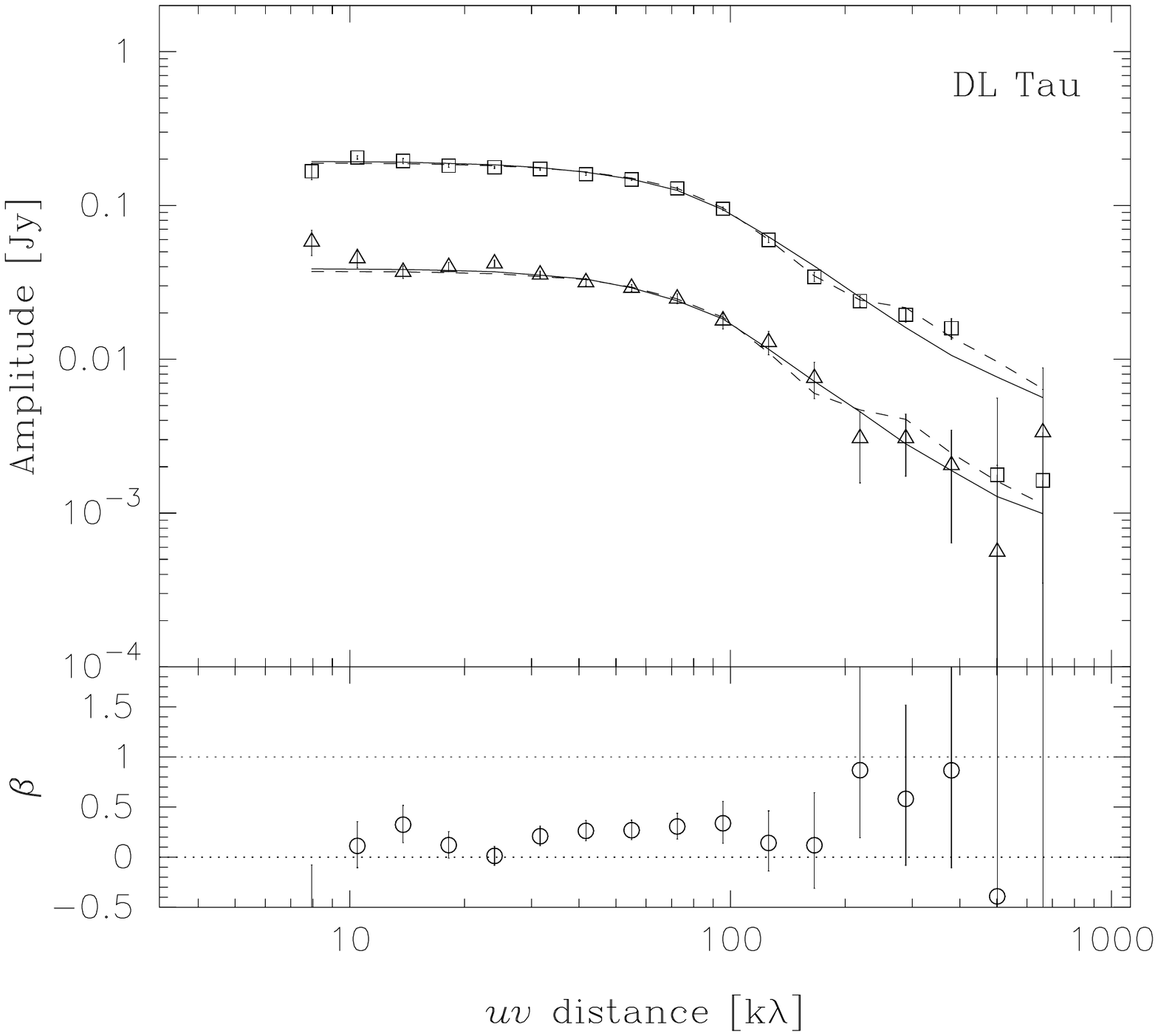} \\
\vspace{-1.0cm}
\includegraphics[angle=270,scale=0.3]{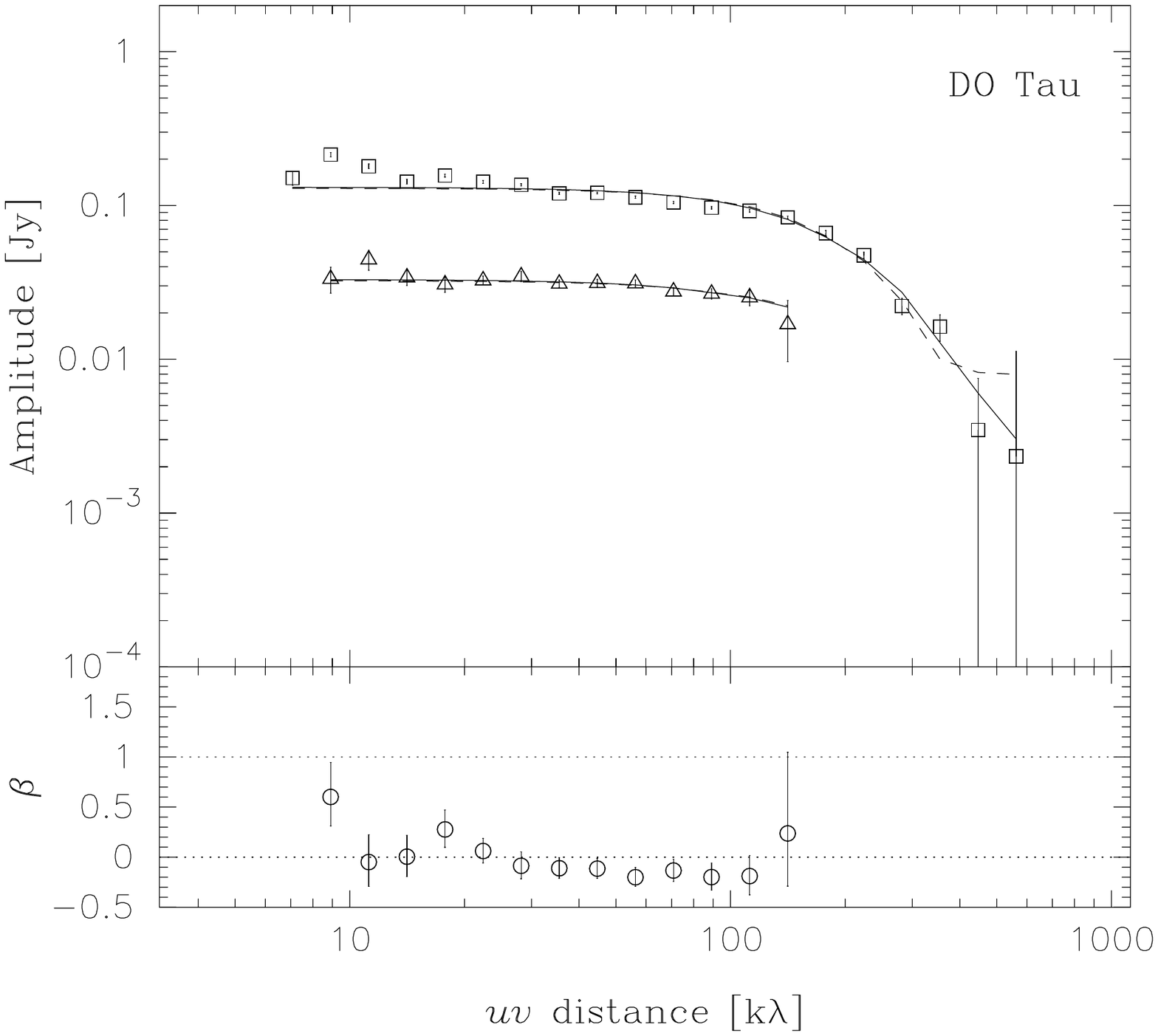} &
\includegraphics[angle=270,scale=0.3]{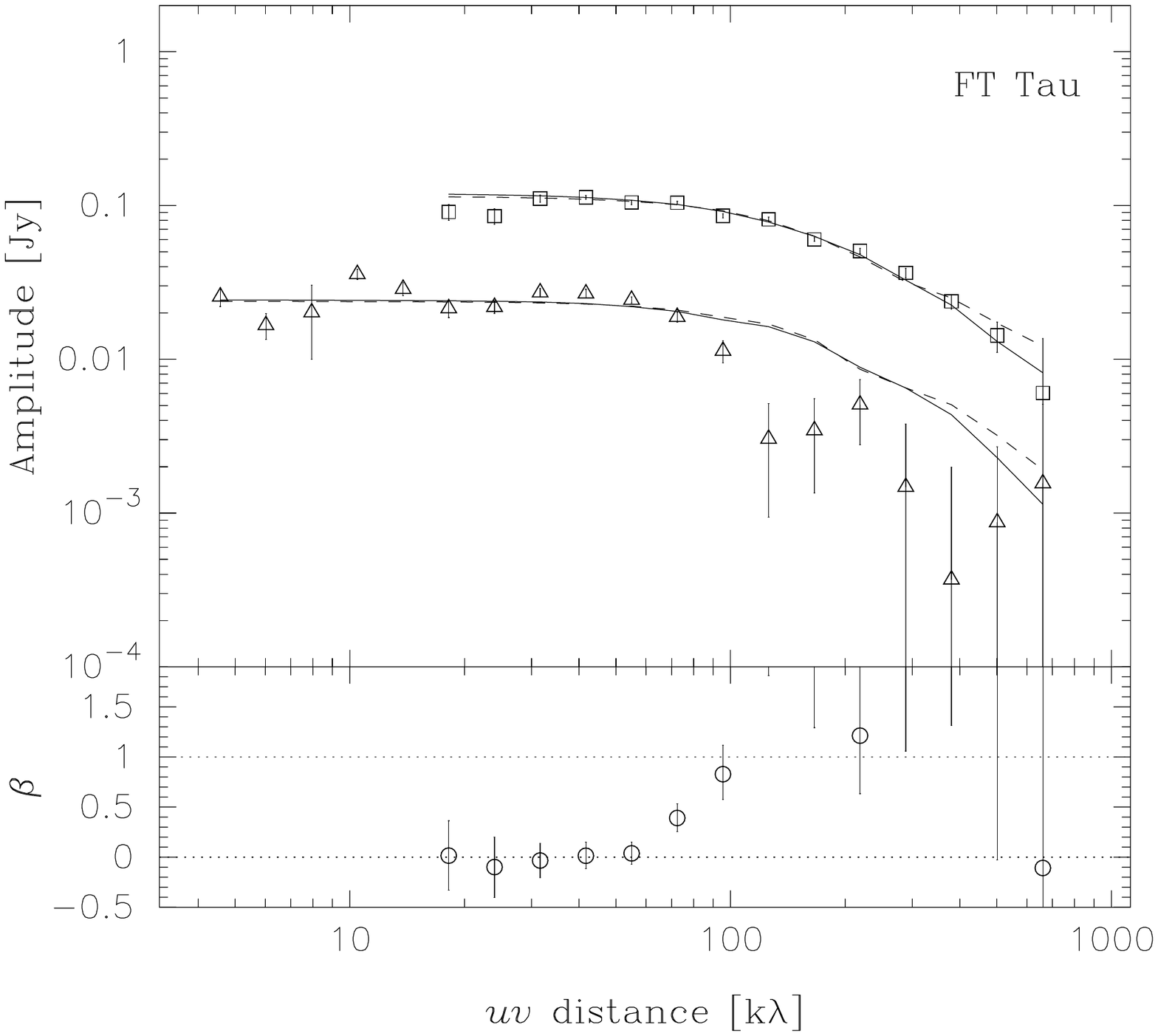} \\
\includegraphics[angle=270,scale=0.3]{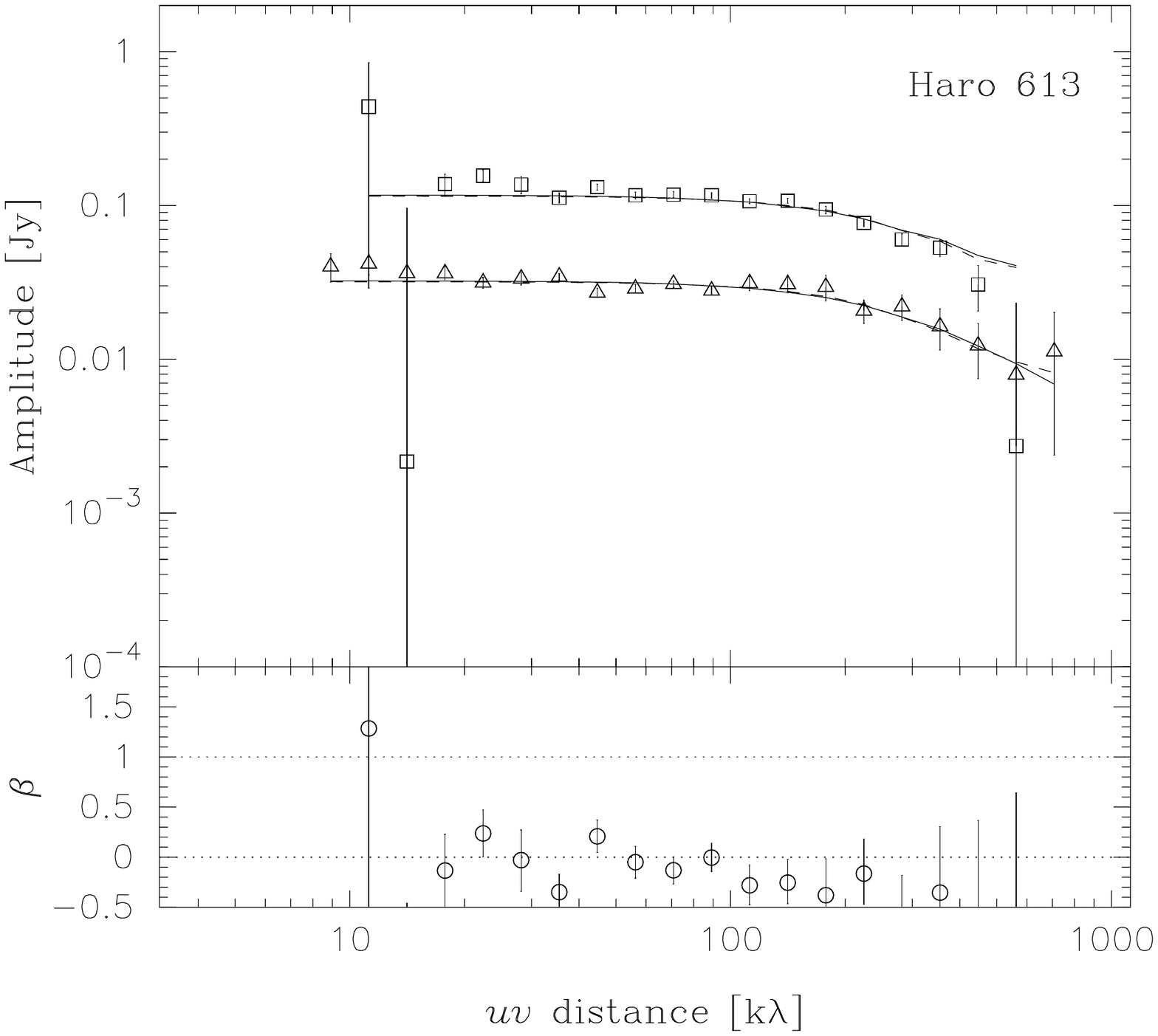} &
\includegraphics[angle=270,scale=0.3]{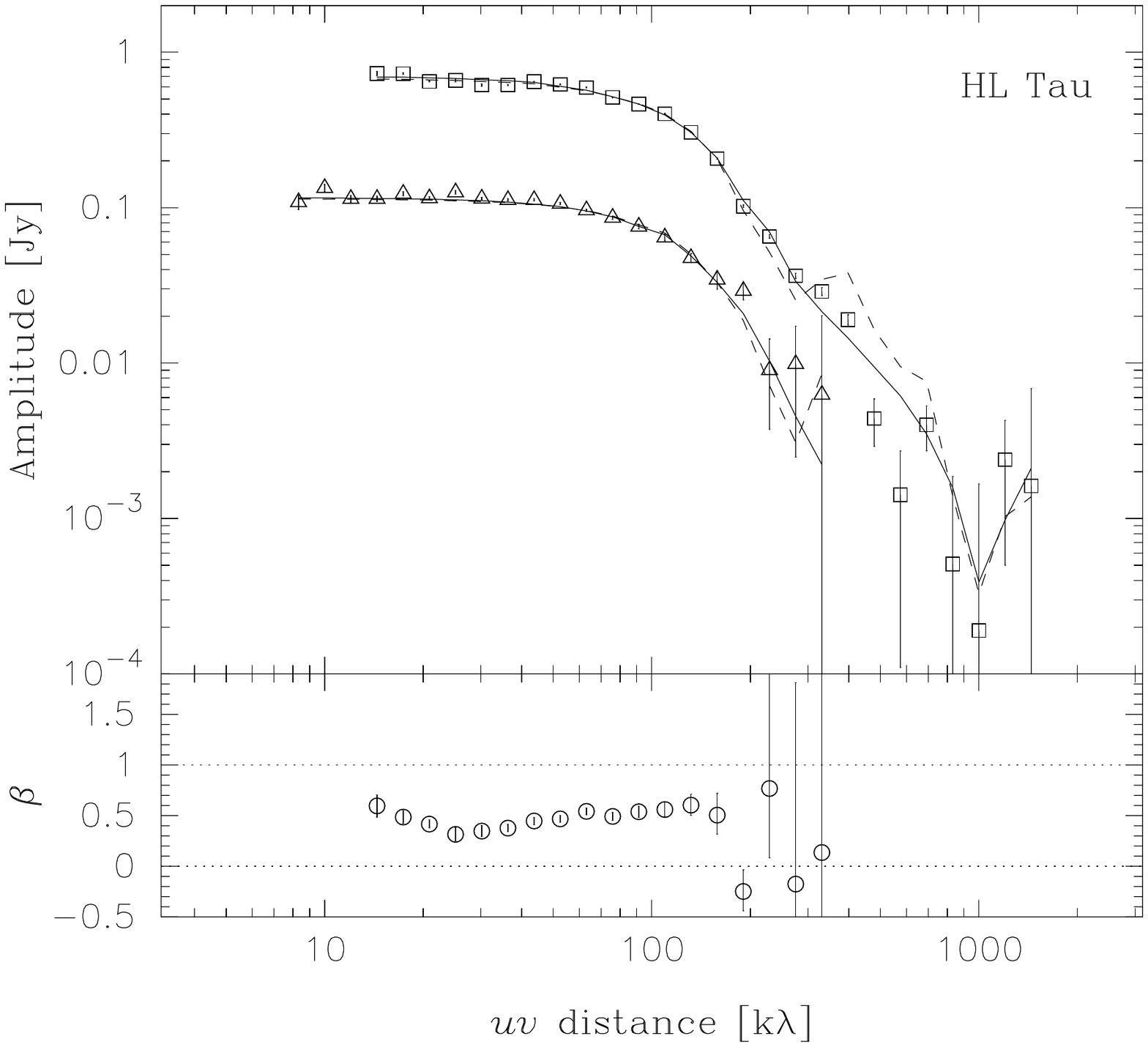} \\
\end{tabular}
\caption[{\it uv} amplitude plots of targets]{
{\it uv} amplitude plots of targets
with their best fitting models for the accretion disk model in solid lines
and the power-law disk model in dashed lines.
$\beta$ values are calculated from the observational
data simply assuming optically thin and Rayleigh-Jeans approximations.
The $\beta$ results without these approximations are shown in the modeling.
Open squares and triangles are for \omm\ and \tmm\ data, respectively.
\label{fig_uvamp}}
\end{center}
\end{figure}

Figure \ref{fig_uvamp} presents observational visibilities of six
targets averaged in annulus with the best fitting model overlaid:
the solid and dashed lines indicating accretion disk and power-law
disk models, respectively. 
The dust opacity spectral indexes at the bottom of each panel
have been calculated in the optically thin assumption and the
Rayleigh-Jeans approximation: $\beta \approx
\textrm{log}(F_{1mm}/F_{3mm})/\textrm{log}(\nu_{1mm}/\nu_{3mm})-2$.
The $\beta$ values from our modeling without such assumption
and approximation are addressed later.
Open squares and open triangles mark $\lambda=1.3$ mm and 2.7 mm
data, respectively.
The error bars are the standard error of the mean:
$\sqrt{\Sigma \sigma_i^2/N^2}$, where
$\sigma_i^2$ is the variance of a visibility data point in a
annulus calculated from its $T_{sys}$ and N is the total number of
data in the annulus.  Absolute flux calibration uncertainty
is not applied for
the plots.  As shown, the two models are fitting observational data
well.  Note that since the viscous accretion disk model does not
have an outer radius, there are no bump features at around
$200-300~k\lambda$, corresponding to about $1''$.

\textbf{Although the $\beta$ values of the figure are simply calculated and not from our modeling, it is worthy of noting the weak relative
variations along {\it uv} distance, which can be used for studying
grain size distributions. In general, recent studies reported that
protoplanetary disks have larger grains in the inner regions
\citep[e.g.,][]{2011A&A...529A.105G,2012ApJ...760L..17P,
2013A&A...558A..64T}.  Observations with higher
angular resolution and sensitivity at multiple wavelengths are
required to investigate the spatial distributions of dust grains.
We do not intend to study the grain size distributions in this paper
and simply assume a constant $\beta$.}

\subsection{Fitting Results}
\label{sec_fittingresults}

\begin{figure}
\begin{center}
\includegraphics[angle=0,scale=0.8]{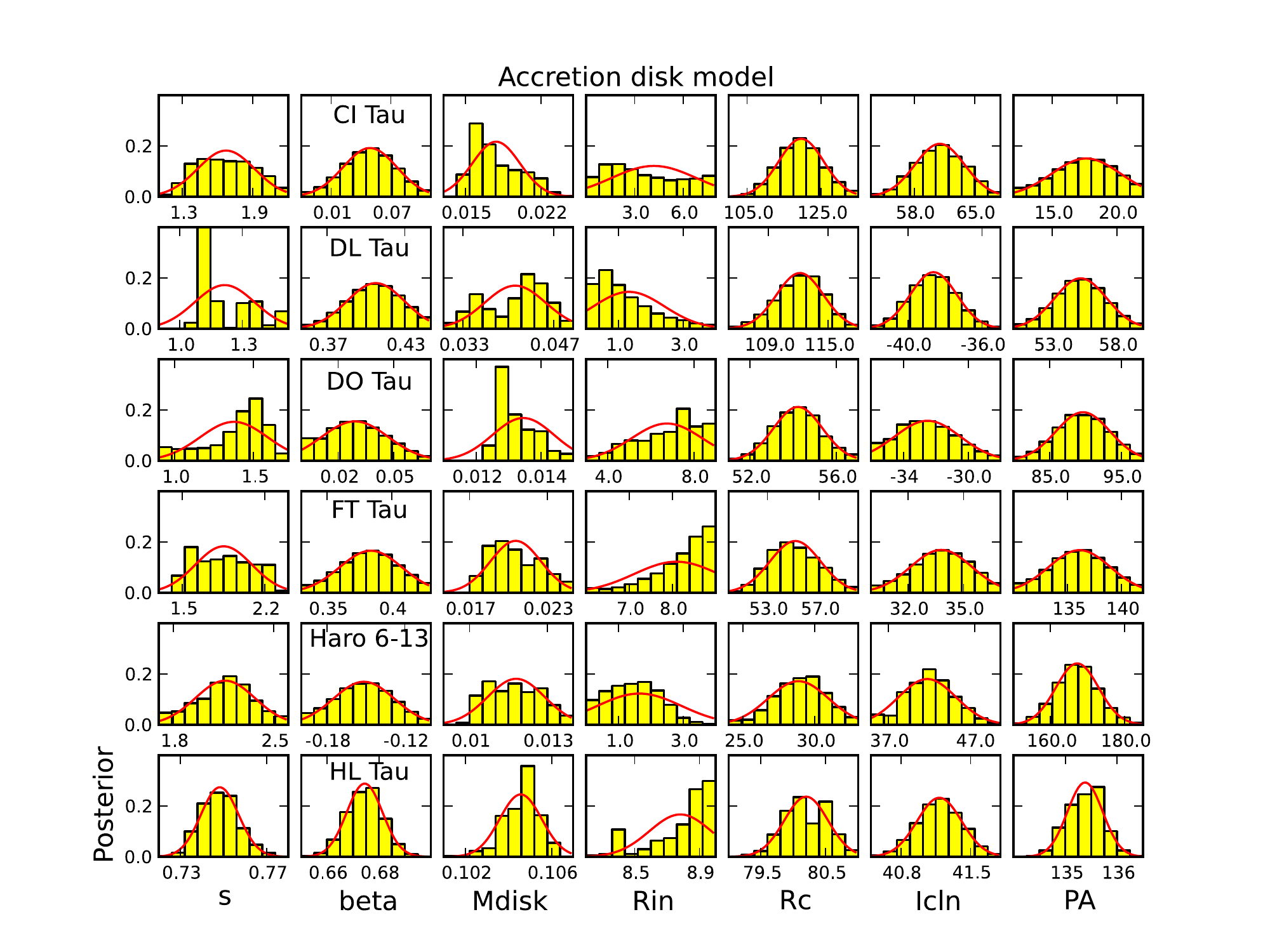}
\caption{Posterior distributions of accretion disk model parameters
fitting 6 disk data. Target names are indicated in the $\beta$ panels,
the second column from the left. The $M_{disk}$ is in units of $M_\sun$,
$R_{in}$ and $R_{out}$ are in AU, and inclination and position angles
are in degree.
\label{fig_postac}}
\end{center}
\end{figure}

\begin{figure}
\begin{center}
\includegraphics[angle=0,scale=0.8]{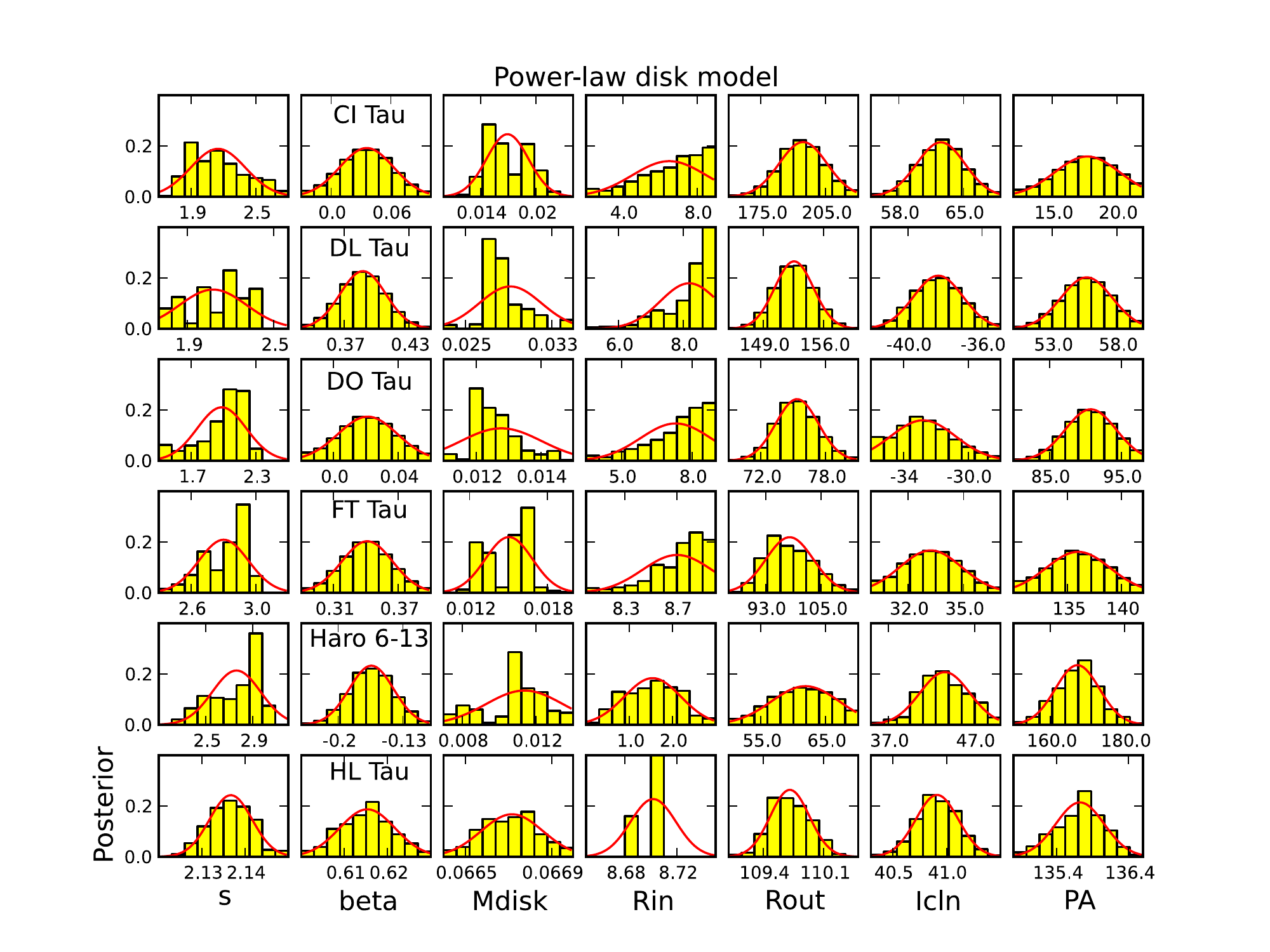}
\caption{Posterior distributions of power-law disk model parameters
fitting 6 disk data.
\label{fig_postpw}}
\end{center}
\end{figure}

We have carried out two flared disk models: viscous accretion disk and
power-law disk models. As described in Section \ref{sec_diskmodel},
we have 7 free parameters and two controlled parameters for the vertical
structures (5 disk thickness $b_t$ and 5 flare index $h$ values).
The fitting results are presented in Table \ref{tab_parampost} and Figures
\ref{fig_postac} and \ref{fig_postpw}.
The 25 cases of the controlled parameters are combined \textbf{in order to see the overall parameter ranges, as they have been sampled with the same Gaussian proposal function (Section \ref{sec_fitting_proc}).}

We did not use any priors obtained from the images of the targets
except the center positions and the inclination signs.
We tested the center positions determined by a Gaussian fit of the IMFIT
task in MIRIAD and the ones manually determined, based on the
intensity peak positions.  The offset values by the simple
Gaussian fit have been chosen (Table \ref{tab_diskdata}), as they give
higher posteriors.

The fitting results are listed in Table \ref{tab_parampost} and the
posterior distributions of parameters in individual targets are
presented in Figure \ref{fig_postac} and \ref{fig_postpw}.
The table values are the means and standard deviations of individual
parameters in the marginal probability distributions.
The solid lines of Figure \ref{fig_postac} to \ref{fig_postpw}
indicate normal distributions.
As shown, most parameters have a normal
distribution posterior, and hence well defined values and deviation.

\textbf{The exception is the inner radius. The inner radius was fit
over a range from 1.3 to 9 AU because none of the disks
showed an inner hole or central depression of emission at our 
best 18 AU resolution. The resulting fits show weak dependence
on inner radius. The statistical uncertainties in Table \ref{tab_parampost} 
do not account for the systematics of increased atmospheric
de-coherence on the longest baseline which are biased toward 
blurring the central disk emission.  The inner radius
for HL Tau, the source with the best resolution and signal-to-noise,
may represent a flattening of the surface density distribution
within 10 AU rather than a central hole
(i.e. ALMA SV data).}

The volume density distribution index ($s$) of the power-law disk
model is between $\sim$2.0 to $\sim$2.8, which corresponds to a
surface density
index ($p$) between $\sim$0.4 and $\sim$1.8 when the best flare
index values are adopted ($p=s-h$ in Section
\ref{sec_diskmodel}).
On the other hand, the $s$ values of the accretion disk
model are in a range of 0.7 to 2.2, which
are converted into $\gamma$ expressing the surface density
distribution ($\gamma=s-h$) of
-0.2 to 1.1, as shown in Table \ref{tab_parampost}.
Such a large range of $\gamma$ has been reported by previous
studies \citep{2009ApJ...701..260I,2009ApJ...700.1502A}.

The dust opacity spectral index $\beta$ appears to
be constrained very closely in the two models.  This is expected because
$\beta$ is sensitive to the difference of flux densities at two
wavelengths.  $\beta$ of our sample disks ranges from -0.15 to 0.70
and half of them are around 0, which can be interpreted as grain growth.
$\beta$ of younger Class 0 YSOs is around 1
\citep{kwon2009}, so grains may grow with object evolution.
This is still valid for DL Tau, FT Tau and HL Tau, since they have $\beta$
smaller than 1. Note that the possible error of $\beta$ due to the
absolute flux calibration uncertainties (10\% at \omm\ and 8\% at
\tmm; refer to Section \ref{sec_obs}) are about 0.25.

Disk masses are 0.01--0.10 \msun.
Compact disks (DO Tau, FT Tau, Haro 6-13) tend to have smaller masses.
However, disk masses are distributed
in a large range, among which HL Tau has the largest mass.  Note that the
disk masses have a large uncertainty mainly because $\kappa_0$ has a factor
of two uncertainty \citep[e.g.,][]{ossenkopf1994}.
In addition, the stellar luminosity affects the disk mass, as it
sets the dust temperature.  However,
disk masses are not so sensitive to the
stellar luminosity due to the low power index and
the high optical depth along radius.  Note that the derived temperature distributions
for our modeling are calculated by the Monte-Carlo radiative transfer
code of RADMC 3D.

The disk sizes (characteristic and outer radii) are well constrained
in both models: $R_c \approx 30$--120 AU and $R_{out} \approx 60$--200
AU.  The characteristic radius of the accretion disk model
is where the disk density distribution
changes from the power-law dominant region to the exponential dominant region
(Equation \ref{eq_accsurfacedensity}).
Comparing the surface density equation of our accretion disk model
with a similarity solution of viscous accretion disks
and considering the mass flow equation
\citep[e.g.,][]{2009ApJ...700.1502A,1998ApJ...495..385H,
1981ARA&A..19..137P},
the characteristic radius can express the transitional radius $R_t$,
where the bulk flow direction changes from inward to outward to conserve
angular momentum \citep[e.g., Equation A9 in][]{2009ApJ...700.1502A}:
\begin{equation}
R_t = R_c\Big[\frac{1}{2(2-\gamma)}\Big]^{1/(2-\gamma)}.
\end{equation}
The $R_t$ of our sample disks are in a range of about 15 to 60 AU, as
listed in Table \ref{tab_parampost}.

The two geometrical parameters, inclination ($\theta_i$) and position
angle (PA), are the two best constrained.  The values are consistent with
previous high angular resolution observations.
In addition, they are consistent in both models, so model-independent.
We originally did not limit the inclination sign based on the
known bipolar outflow directions of our targets in the modeling.
However, for the final modeling results, we have chosen the inclination
sign in agreement with the bipolar outflow direction and/or
previous studies \citep[e.g.,][]{2011A&A...529A.105G}.

\textbf{Five of our disks were also observed and fit to power-law and accretion disk
models by \citet{2011A&A...529A.105G}. Disk inclination and position
angles agree fairly well (accounting for the difference in definition of position angle) but
other disk parameters show only broadly similar trends, without detailed agreement. For
HL Tau and Haro 6-13, our angular resolution, which is a factor of 3-4 higher, is the likely
reason for the difference. For CI Tau, where the resolutions of the observations are
nearly the same, the fitted disk parameters, $R_c$, $\gamma$, $R_{out}$, and $p$,
are similar, but generally outside of the statistical uncertainties by more than 2$\sigma$.
 \citet{2011A&A...529A.105G} assume a simple $R^{-0.4}$ power-law for the radial temperature variation which may be responsible in part for the difference. However,
it is likely that both analysis are more limited by the sensitivity and angular resolution
of the observations than is reflected in the statistical uncertainties. This is a cautionary
note for all disk modeling to date.}

\subsection{Model Comparison}

First, we compared posteriors of the two models.
Bayesian inference provides a way for model comparison;
models can be compared by evidence, which
is the integration of likelihood times prior over the whole possible parameter
space of a model \citep[e.g.,][]{mackay2003}.
We set the prior as well as the number of free parameters
the same for the two models (uniform priors over
the same ranges),
so we just need to integrate likelihood.
The likelihood integration can be estimated
by the best likelihood times the posterior
accessible volume:
\begin{eqnarray}
K &=& \frac{P(D\mid H_A)}{P(D\mid H_P)} \nonumber \\ &=& \frac{\int dm P(D\mid m,H_A) P(m\mid H_A)}{\int dm
P(D\mid m,H_P) P(m\mid H_P)} \nonumber \\
 &\approx& \frac{P(D\mid m_{best},H_A) P(m_{best}\mid H_A) \sigma_{(m\mid
D)_A}}{P(D\mid m_{best},H_P) P(m_{best}\mid H_P) \sigma_{(m\mid D)_P}} \nonumber \\
 &\approx& \frac{P(D\mid m_{best},H_A) \sigma_{(m\mid
D)_A}}{P(D\mid m_{best},H_P) \sigma_{(m\mid D)_P}},
\end{eqnarray}
where the subscript $A$ and $P$ indicate the accretion disk model and
the power-law disk model, respectively.
The posterior accessible volume is estimated by $\sigma_{(m\mid D)}
 = \Pi \sigma_i$, where $\sigma_i$ is posterior widths of
$s$, $\beta$, $M_{disk}$, $R_{in}$, $R_{out}$ (or $R_c$),
$\theta_i$, and $PA$.

The K values of our targets are shown in Table \ref{tab_modelcomp}
with the evidence of the two models.  As shown in the table, all
the targets, except DL Tau, have a positive value in $\textrm{ln}(K)$.
Particularly, HL Tau, whose data have the highest angular resolution
in our sample, has a significantly larger positive value.
A scale for interpretation of these values has been proposed:
$K > 3$ indicates a considerable preference for the model of the
numerator (here the accretion disk model) and $K < 1/3$ for the
denominator model (here the power-law disk model)
\citep[e.g.,][]{Kass:1995vb}.  Therefore, we conclude that overall
the accretion disk model is preferred.  It is noteworthy that the
preference of the accretion disk model is quantitative and purely
based on continuum data.  \citet{2008ApJ...678.1119H} qualitatively
argued that the accretion disk model with a tapered outer region
instead of the power-law disk model with a sharp outer radius is
preferred, as it explains more extended disk features detected in
gas tracers.  A previous work on two protoplanetary disks reported
slightly smaller reduced $\chi^2$ values for the accretion disk
model than the power-law disk model \citep{2010ApJ...714.1746I}.

In contrast to the overall trend, DL Tau prefers the power-law disk
model, which may be implying a difference in evolutionary state or
physical conditions in the disk.  In addition, the model preference
is not significant toward CI Tau and Haro 6-13.  It may be due to
the non-axisymmetric feature in CI Tau and the small disk size of
Haro 6-13. Note that Haro 6-13 has the smallest disk in our sample,
slightly larger than our angular resolution.

\textbf{\citet{2011A&A...529A.105G} also reported no overall preference between
the two disk models in observed disks with sizes comparable to the angular resolution.
However, they found that HL Tau prefers a power-law disk model, while
DL Tau prefers an accretion disk model, unlike our results.
Although their data have slightly lower angular resolution than ours, they
took the radial dependence of $\beta$ into account. We have only
a constant $\beta$, but we employ a better temperature distribution and a
vertical structure. The discrepancy simply presents the difficulties of model
comparison using the current data sets.}

In fact, disk model comparison
studies may require higher sensitivity and resolution data for a
larger sample, as well as detailed modeling, in order to achieve a reliable result.
We do not attempt further
discussion of the model comparison results in this paper.  Instead, we
further investigate the disk properties obtained only from the
accretion disk model in the following.

\subsection{Residual Maps}

\begin{figure}
\begin{center}
\begin{tabular}{c@{\hspace{-0.7cm}}c}
\vspace{-1.0cm}
\includegraphics[angle=270,scale=0.3]{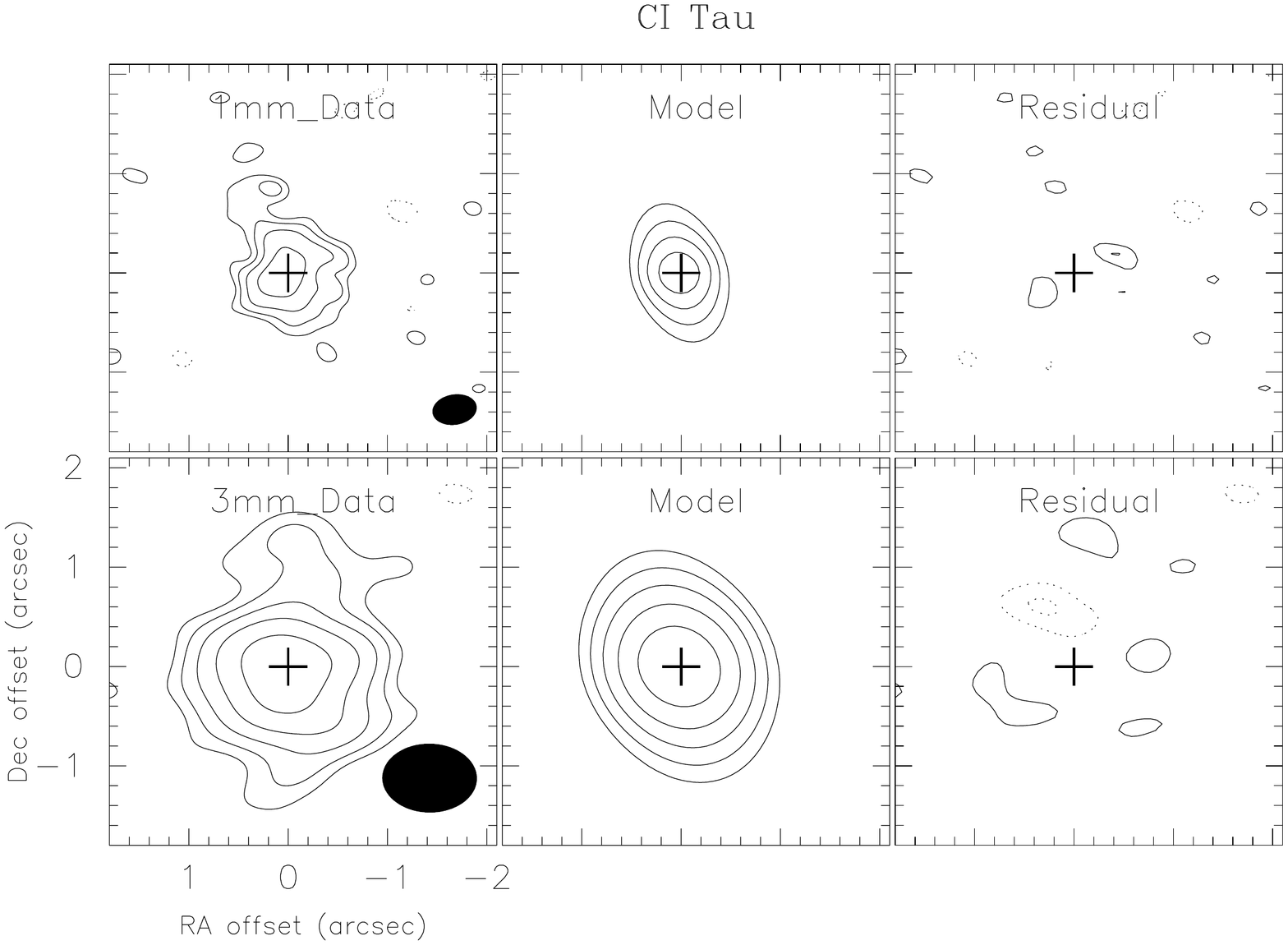} &
\includegraphics[angle=270,scale=0.3]{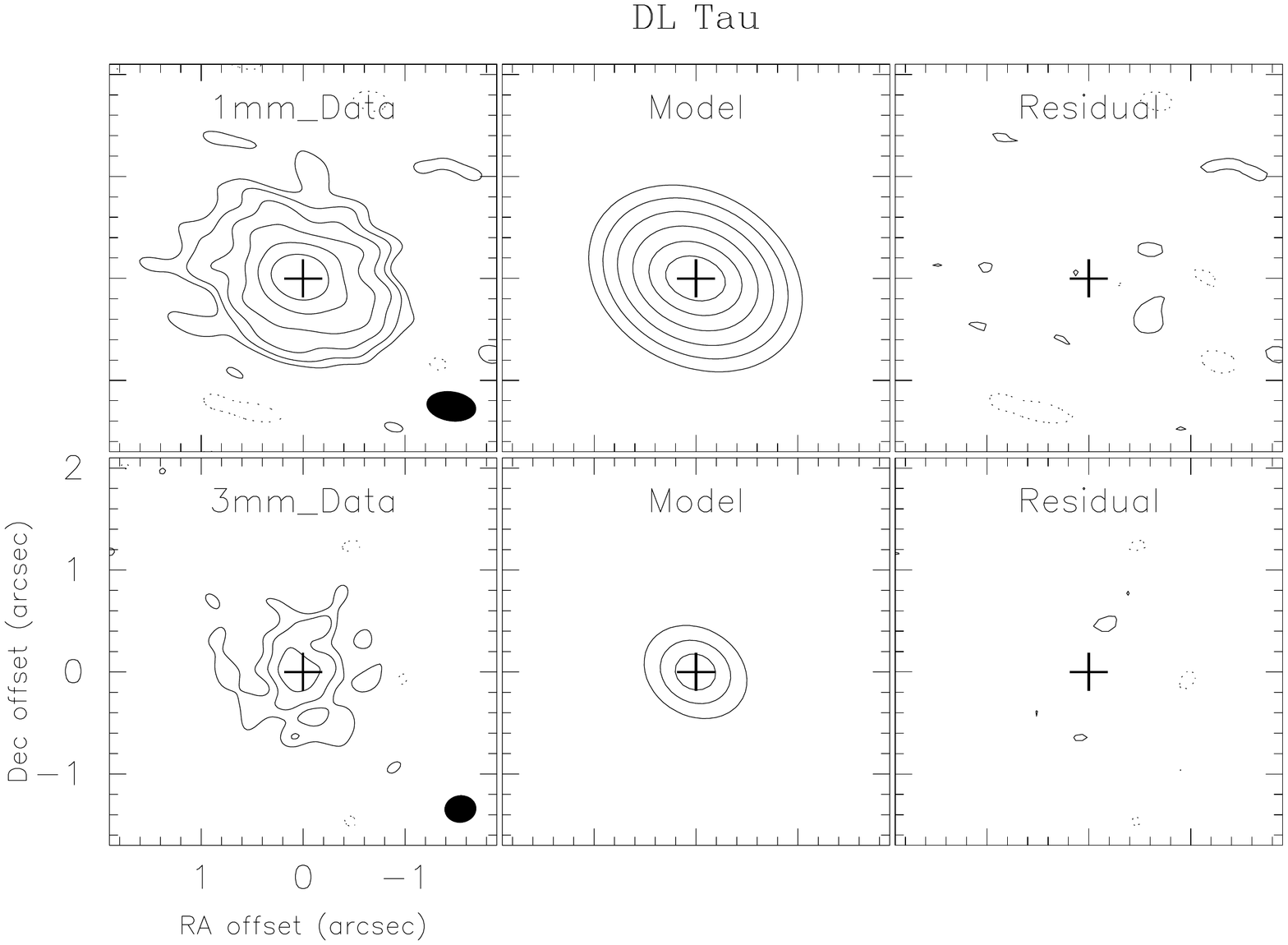} \\
\vspace{-1.0cm}
\includegraphics[angle=270,scale=0.3]{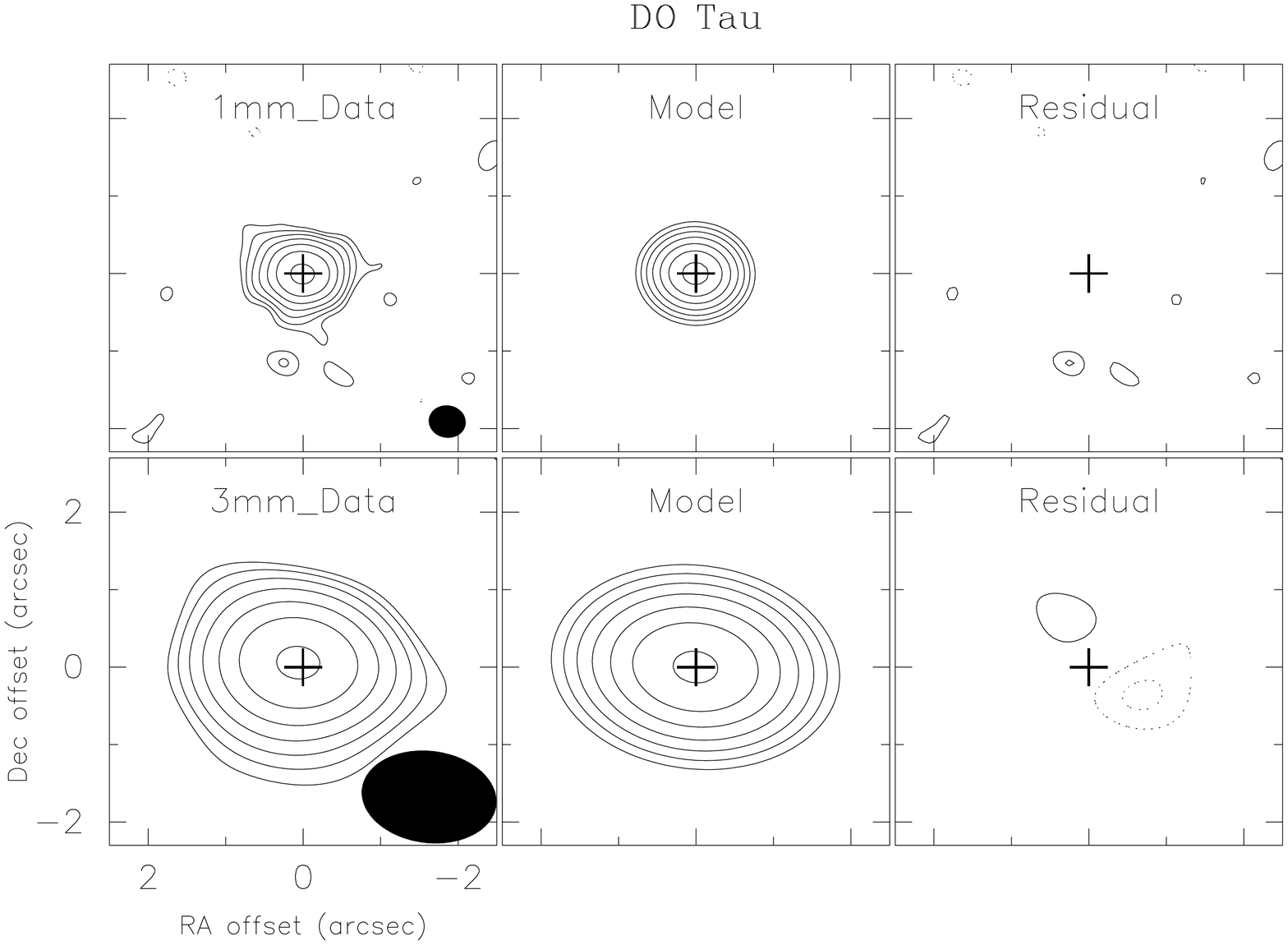} &
\includegraphics[angle=270,scale=0.3]{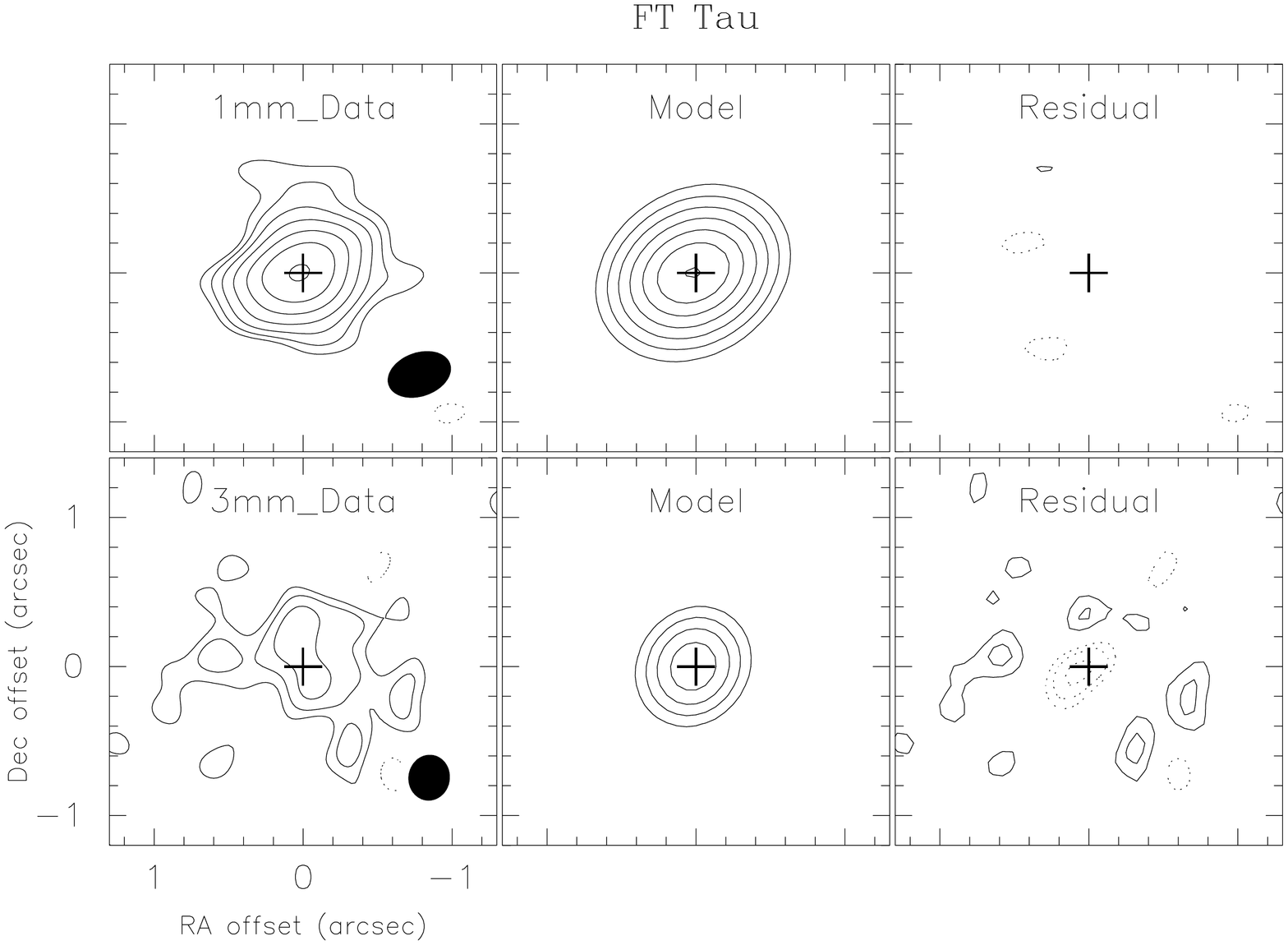} \\
\includegraphics[angle=270,scale=0.3]{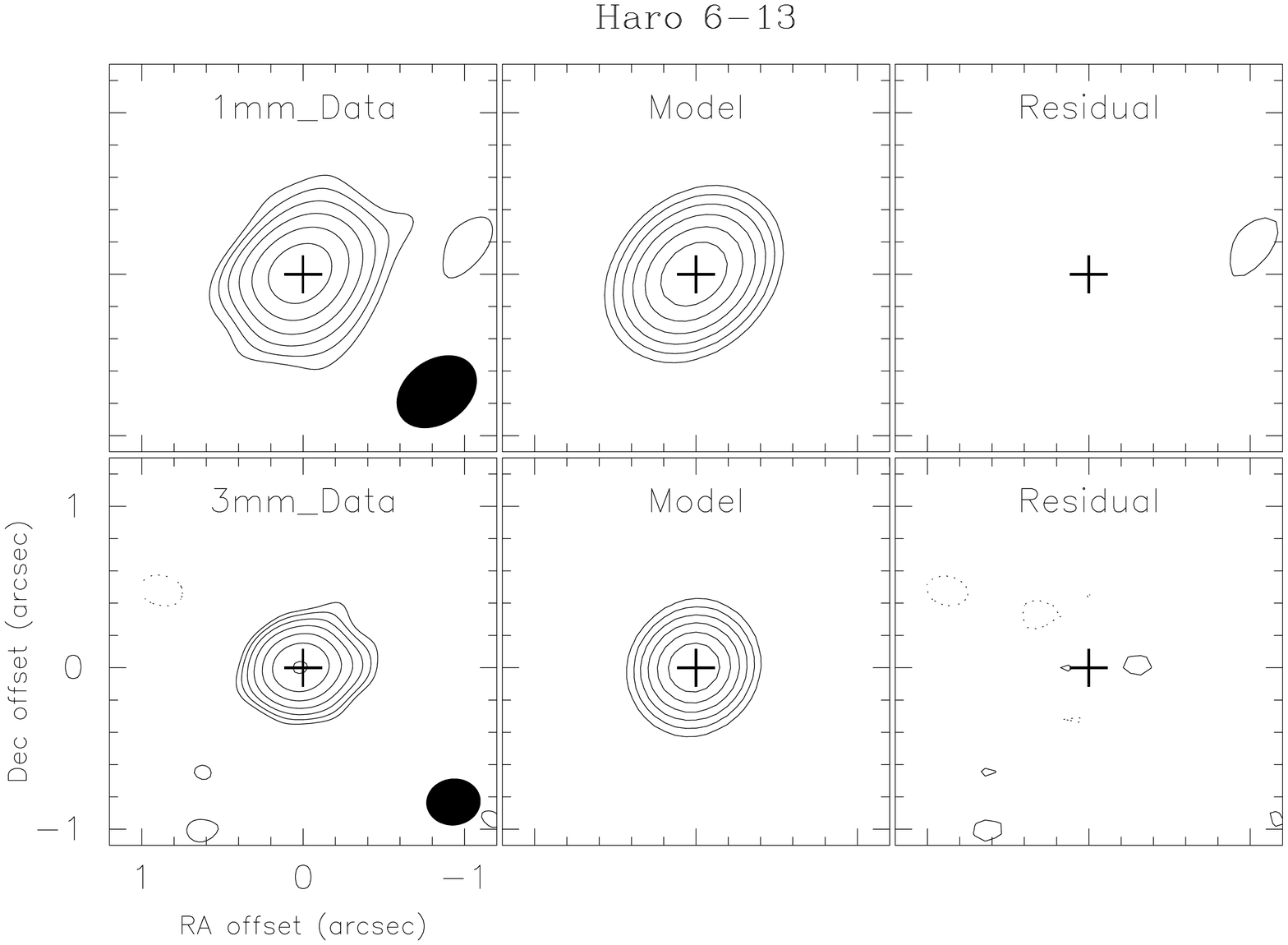} &
\includegraphics[angle=270,scale=0.3]{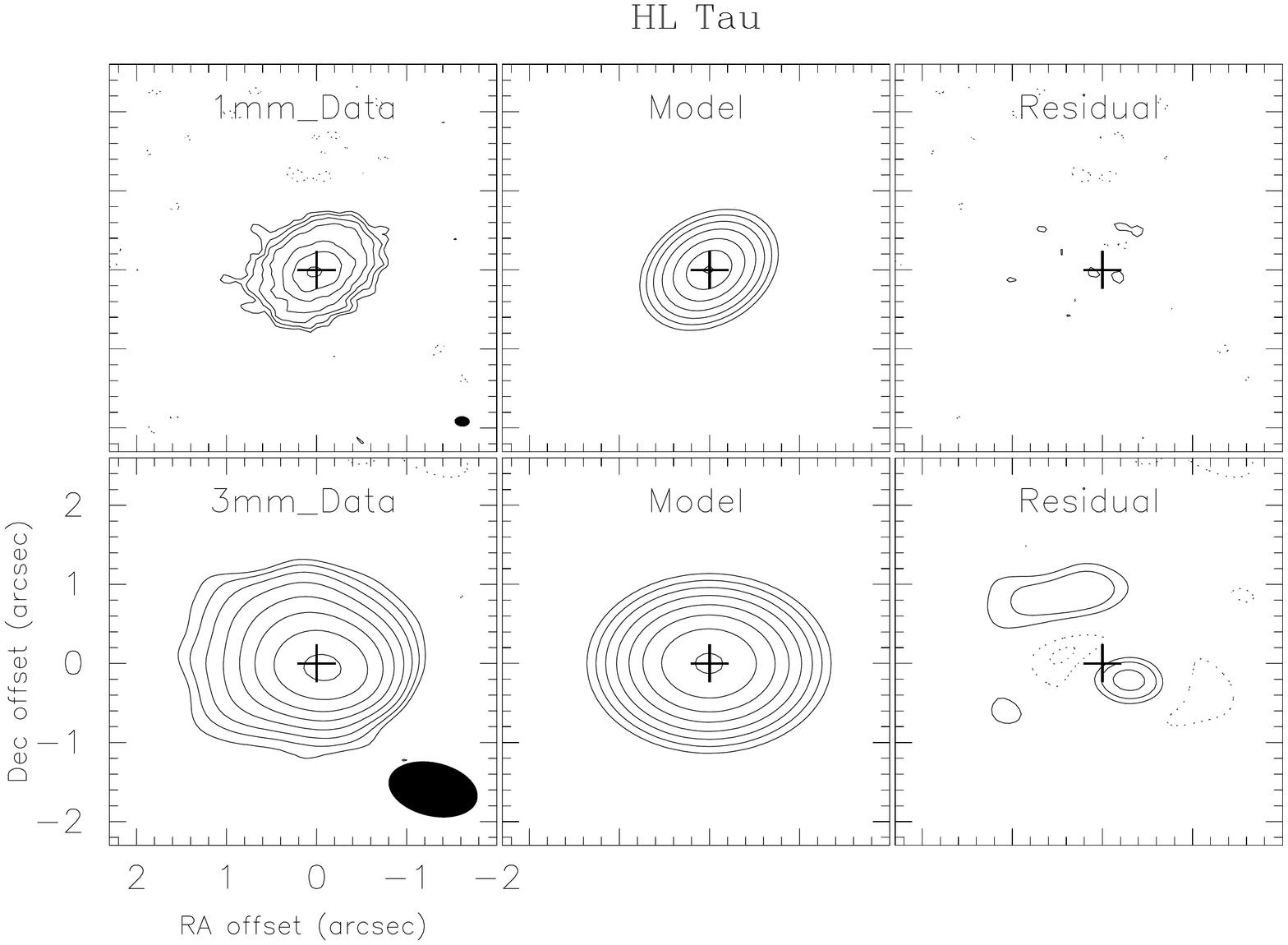} \\
\end{tabular}
\caption
{Disk continuum, model, and residual maps at \omm\ and 2.7 mm for
the accretion disk models.  From the left, individual target
observational images, best models, and residual maps, and the upper
and lower panels are at \omm\ and 2.7 mm, respectively.  Note that
the \omm\ images are the same as the color images of Figure
\ref{fig_diskimages}.  The images are re-centered by the phase-center
offsets listed in Table \ref{tab_diskdata}, and the contour levels
are $\pm$ 2.5, 4.0, 6.3, 10, 16, 25, 40, and 100 times individual
target $\sigma$ listed in Table \ref{tab_diskdata}.  The synthesized
beams are marked in the bottom right corner of the observational
images.
\label{fig_OMR}}
\end{center}
\end{figure}

Figure \ref{fig_OMR} presents the residual maps for the best accretion
disk models with the \omm\ and 2.7 mm continuum images: observational
images, best models, and residual maps from the left.
The residual maps do not have any significant features
except FT Tau and HL Tau.  The scattered peaks of FT Tau at \tmm\
are mainly due to the noisy data of the longest baselines taken in
A-configuration.  The residual features of HL Tau have been discussed
in \citet{2011ApJ...741....3K}: the \tmm\ features may be due to
free-free emission contamination and the \omm\ features are
hints of substructures now seen in the ALMA SV data.  The low levels of the residual indicate
excellence of our model fitting.

Although the residual maps illustrate that the data are consistent
with our models, the well-resolved map of CI Tau has a weak substructure
that is seen at both wavelengths;
a non-axisymmetric
feature is present from north-north-east to south-south-west
at \omm\ and from north to south at \tmm.  Although the feature
is weak, the similar pattern at both wavelength data sets provides a
better fidelity for a non-axisymmetric substructure.
DL Tau and HL Tau do not show a distinct feature for a substructure,
but the extended disks include a possible region for the gravitational
instability.
We discuss the possibility of substructure development on these
disks with the Toomre Q parameter later.  The HL Tau data and images
have been presented in \citet{2011ApJ...741....3K}, but we have
employed a slightly different accretion disk model with an independent
disk flareness parameter and a more realistic temperature distribution
in this paper and carried out modeling identically for all the six
targets.

\subsection{Correlations among Properties}
\begin{figure}
\begin{center}
\begin{tabular}{c@{\hspace{-0.3cm}}c}
\includegraphics[angle=0,scale=0.4]{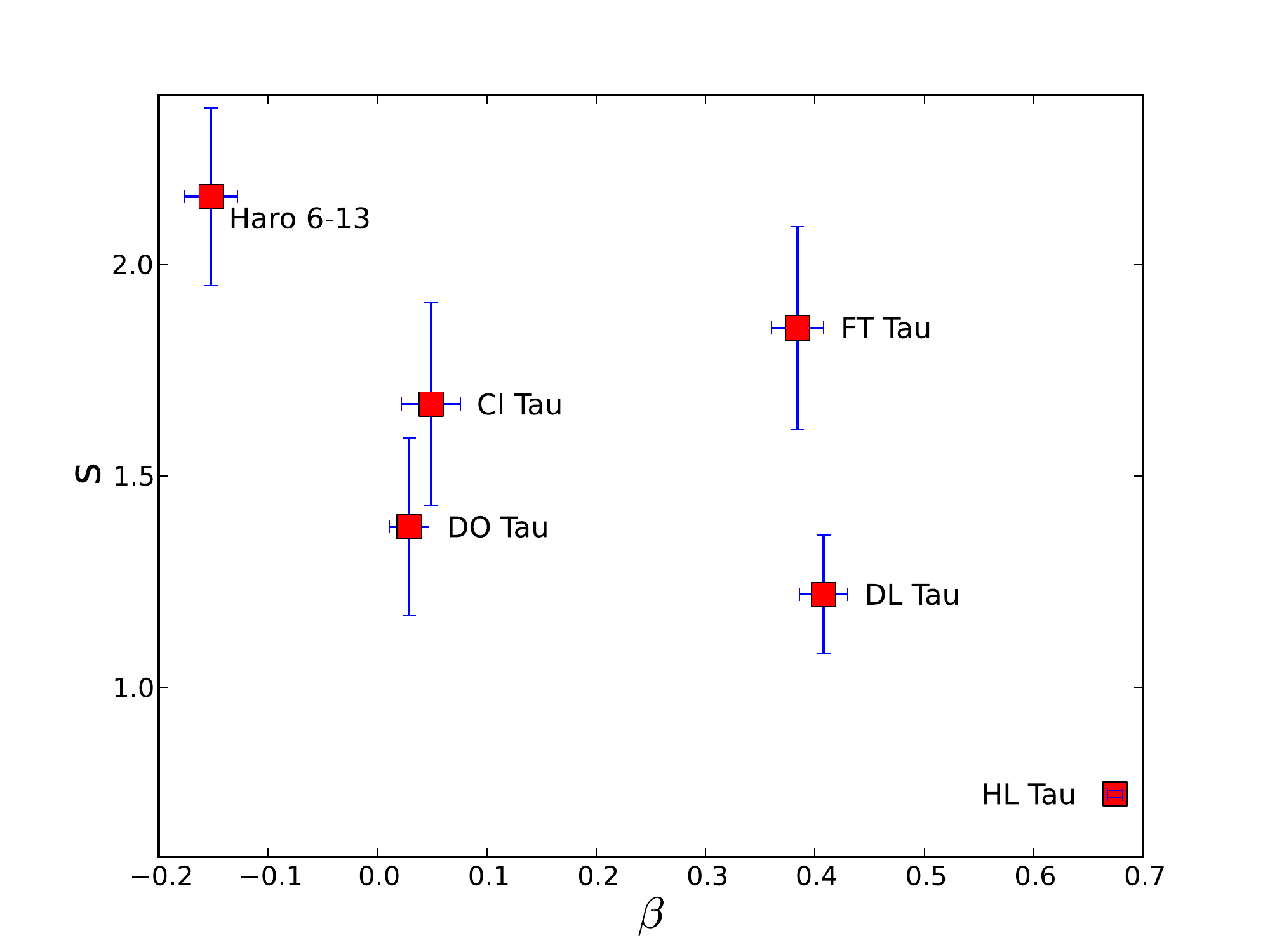} &
\includegraphics[angle=0,scale=0.4]{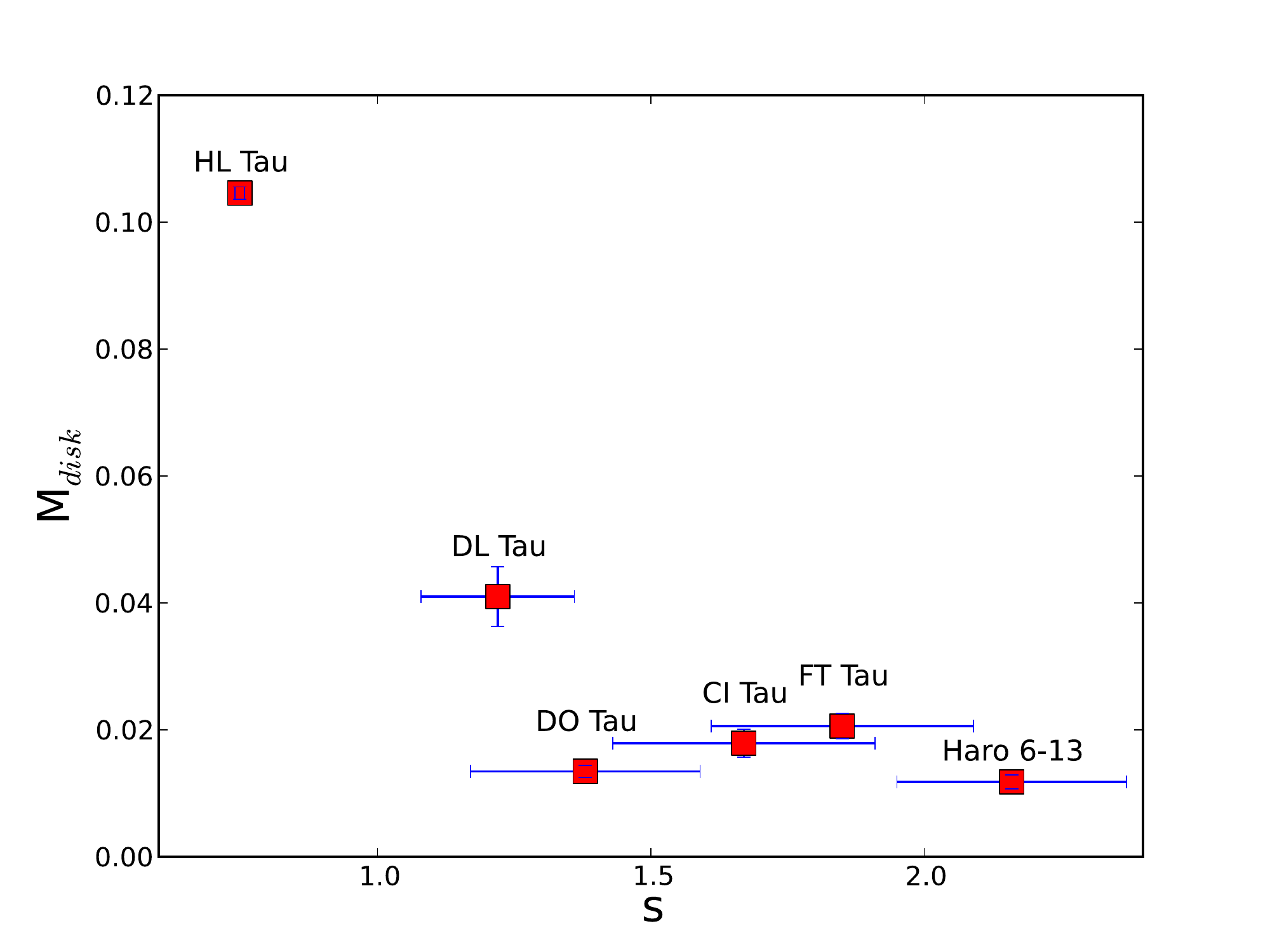} \\
\vspace{-0.3cm}
\includegraphics[angle=0,scale=0.4]{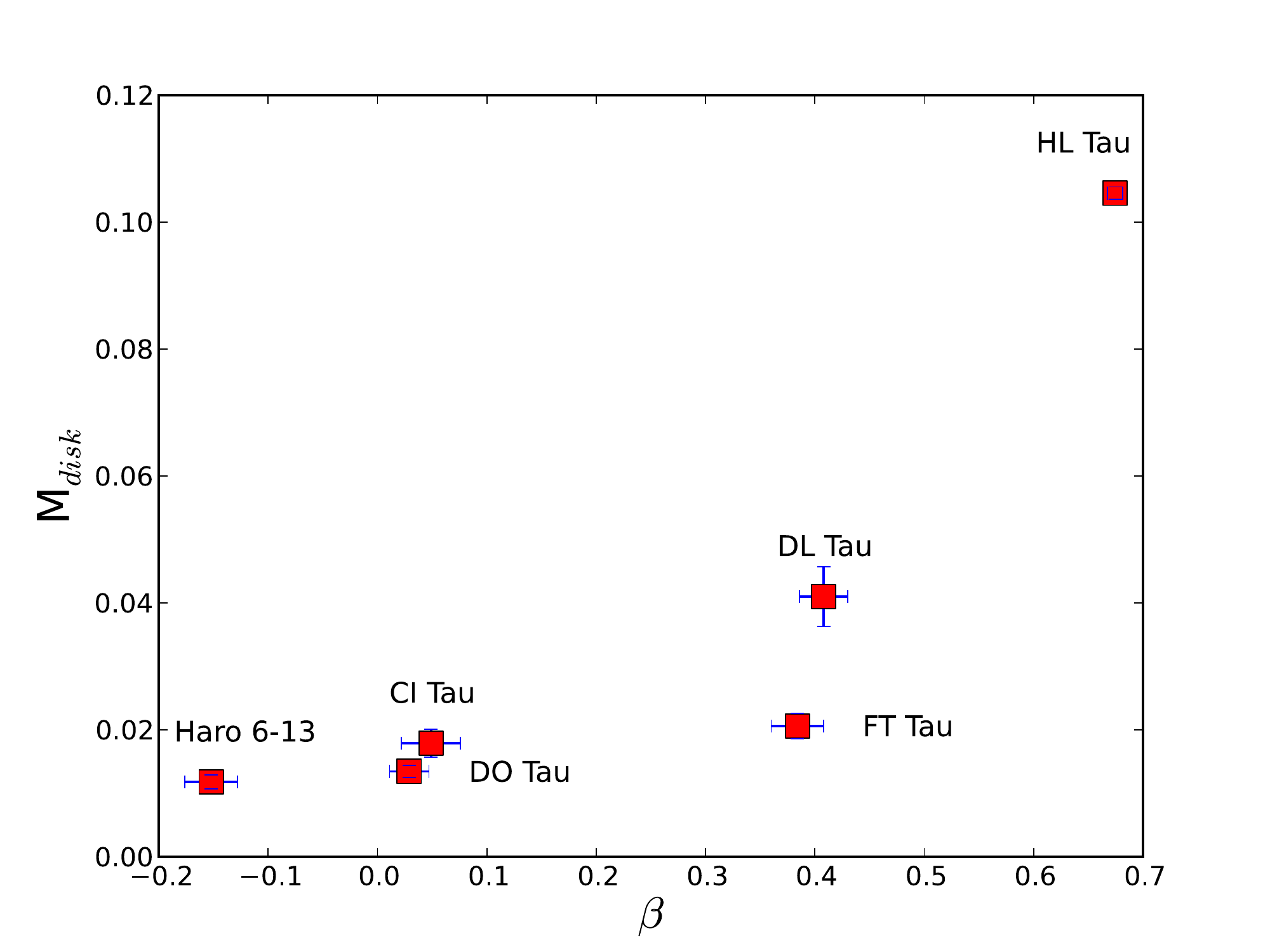} &
\includegraphics[angle=0,scale=0.4]{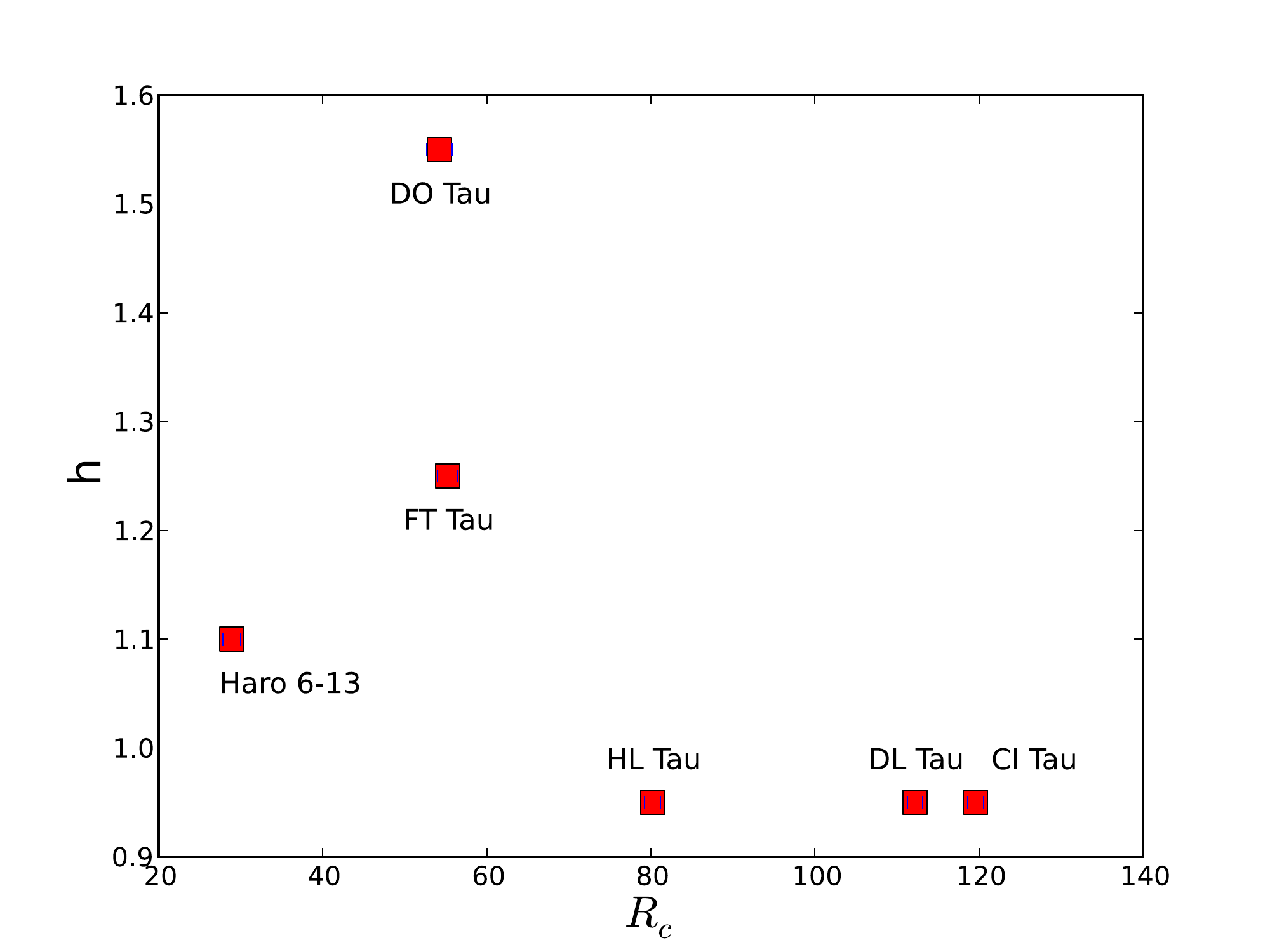}\\
\end{tabular}
\caption{Relationship between properties: the mid-plane density
gradient $s$ vs. $\beta$,
$M_{disk}$ vs. $s$, the flare index $h$ vs. $R_c$, and $M_{disk}$ vs. $\beta$
from the upper left clockwise.
\label{fig_correl}}
\end{center}
\end{figure}

We calculated correlations between the constrained physical properties using
\begin{equation}
R_{xy}=\frac{\Sigma(x_i-\bar{x})(y_i-\bar{y})}
{\sqrt{\Sigma(x_i-\bar{x})^2} \sqrt{\Sigma(y_i-\bar{y})^2}},
\end{equation}
where $\bar{x}$ and $\bar{y}$ are means of $\{x_i\}$ and $\{y_i\}$,
respectively.
The correlation coefficients are presented in Table \ref{tab_correl}.
Relatively strong correlations ($>0.5$) are emphasized with a bold
font and plotted in Figure \ref{fig_correl}.  Although these
correlations are obtained in our small sample of six targets,
a few relationships may provide meaningful
hints on disk properties and evolution.
\textbf{In addition, note that the four correlations are the strongest ($>0.5$)
even without HL Tau data points that look dominant in the relationships.}

\label{sec_beta_diskmass}
As we obtained multiple wavelength continuum data and fit the data
simultaneously with a single model, we can estimate the
dust opacity spectral index ($\beta$).
Overall, $\beta$ of our sample protoplanetary disks are
0.6 or less, which indicates that grains have grown larger
than the younger protostellar systems \citep{kwon2009},
as mentioned in Section \ref{sec_fittingresults}.
Previous studies using high angular resolution data over a broad
wavelength region (submillimeter to centimeter) have reported radial
dependence of $\beta$ \citep[e.g.,][]{2012ApJ...760L..17P}.
\textbf{Our model fitting does not show evidence for such a radial dependence in
$\beta$. In specific, our sources with the most complete data out
to large {\it uv} distance at 1.3 and \tmm, DL Tau, Haro 6-13, and
HL Tau, have no radially symmetric residuals in Figure 5. This is
in conflict with \citet{2011A&A...529A.105G} which finds significant
radial variation in $\beta$ which should have been seen in our data.
The  $\beta$ versus radius for our best resolved disk, HL Tau, is
not given in \citet{2011A&A...529A.105G}. Higher resolution and
high sensitivity observations over a larger range of wavelengths are
needed to solve this discrepancy.}

We found a correlation between $\beta$ and the disk masses:
disks with a smaller $\beta$ are less massive.
Small $\beta$ indicates large grains.
Therefore, this relationship implies that less massive disks have a
larger fraction of large grains.
One caveat on this correlation is that $\beta$ close to 0 may result in
an underestimate of the disk mass because $\kappa$ could be
significantly different from the assumed value.  As an extreme, 
$\beta=0$ is consistent with any size dust grains larger than roughly
an observational wavelength -- pebbles, rocks, etc.  This relationship
has also been found in circumstellar disk around the more massive
young Herbig AeBe stars \citep{2011ApJ...727...26S}.

We also found that the mid-plane density ($s$) is anti-correlated
with the opacity spectral index ($\beta$) and the disk mass
($M_{disk}$). In other words, protoplanetary disks with smaller
$\beta$ and of a smaller mass have a steeper mid-plane density
gradient.
\textbf{This indicates that disks with large grains tend to be 
detected with lower mass and to have relatively larger mass
in the inner regions.}
The increase of the mid-plane density gradient suggests that
large grains move radially as well as settle down in the mid-plane.
However, $\beta$ does not have a strong relationship with the flare
index $h$, which implies that grain growth can occur independently
of disk flatness and grain settlement.
It is noteworthy that there are
other mechanisms of grain growth and disk flatness such as grain
drift and disk evaporation.
\textbf{On the other hand, grain settlement is largely affected by turbulence, which
can be studied by disk gas tracers.}
Interestingly, the flare index does have a relationship with disk
size as shown in Figure \ref{fig_correl}
and described in Section \ref{sec_vstructure}.

However, we did not obtain anti/correlations supporting the ideas
that the equivalence width of small grains (EW) and the spectral
index between 13 and 31 \micron\ ($n_{13-31}$) decrease when grains grow
and disks flatten.  Conversely, we obtain the opposite trends
(see Table \ref{tab_correl}).
For example, $\beta$ is anti-correlated with
EW, which implies that disks with large grains
(a smaller $\beta$) have more fine grains (a larger EW).
In the sense that our millimeter wavelength data are sensitive
to large grains and infrared data are sensitive to fine grains,
it can be interpreted with a more dusty disk.
In addition, note that disks can be detected as a thick disk
at short infrared wavelengths and as a thin disk at long
millimeter wavelengths: stratified grain settlement
\citep{2011ApJ...741....3K}.

\subsection{Disk Vertical Structures}
\label{sec_vstructure}
\begin{figure}
\begin{center}
\includegraphics[angle=0,scale=0.8]{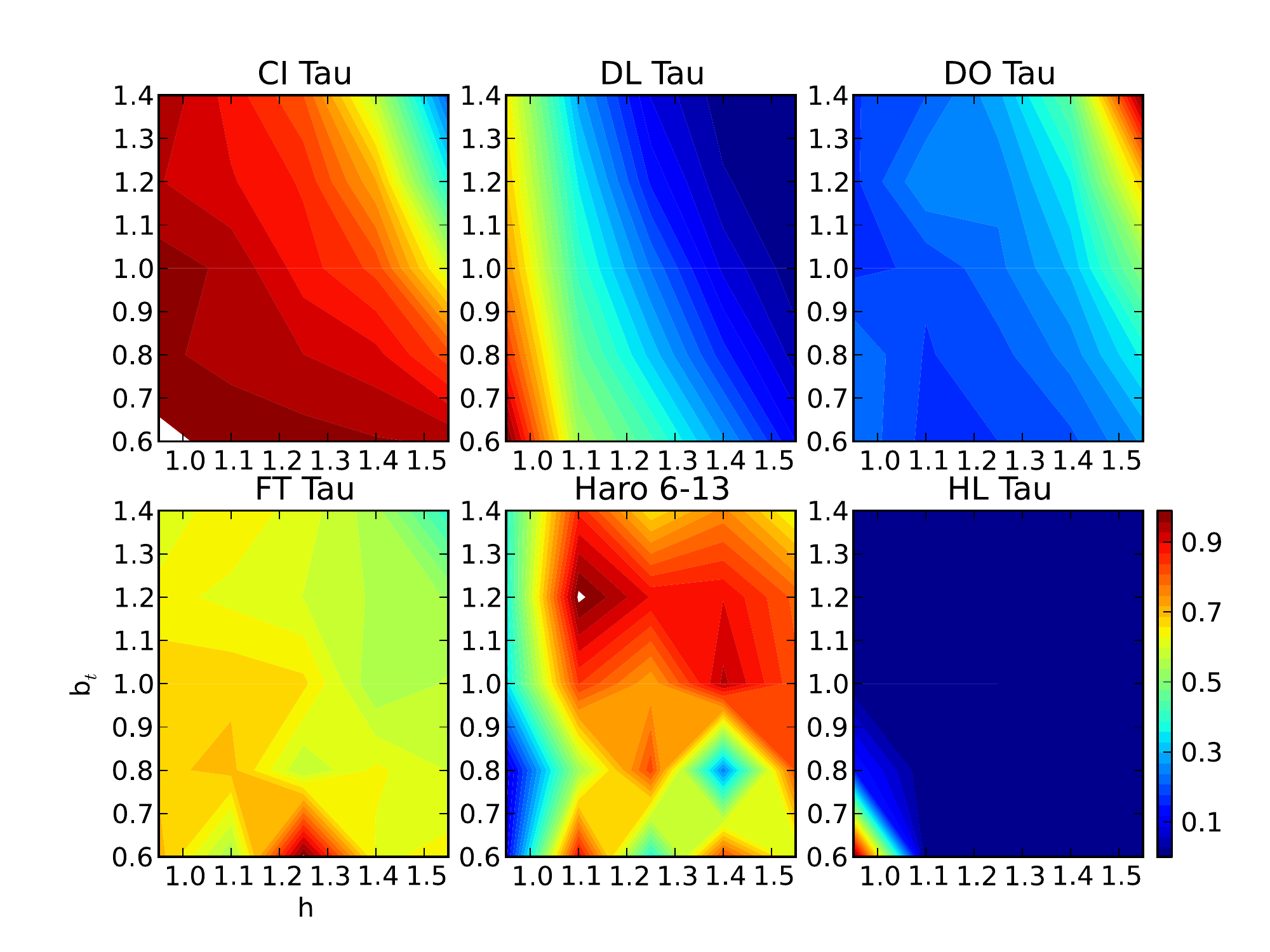}
\caption[Disk vertical structure parameter plots]
{Disk vertical structure parameter plots. The posterior is normalized by
the peak and red indicates high posterior regions.
\label{fig_bh}}
\end{center}
\end{figure}

Figure \ref{fig_bh} shows the posterior distributions of the accretion
disk models in the disk thickness (b$_t$) and the flareness index
(h), normalized by individual peak values.
As shown, the three extended disks (CI Tau, DL Tau, and HL Tau) have
higher posteriors in thinner and less flared models. In contrast, DO Tau
appears to be better fit by thick and more flared models.
In the cases of FT Tau and Haro 6-13, distinct preferences are not
present.  We argue that the sensitivity and angular resolution of the
data are not good enough for these objects.
Note that they are the two faintest targets and compact disks of
our sample.

While our modeling results of millimeter wavelength data
prefer thinner and less flared disks
for CI Tau, DL Tau, and HL Tau, a thick disk model can be required
for shorter mid-IR observations.
For example, HL Tau needs a thick disk for its mid-IR fluxes,
as shown in \citet{2011ApJ...741....3K}. The discrepancy can be understood
by stratified grain settlement: large grains, which millimeter
data are sensitive to, have been settled down in the mid-plane,
while fine grains are not.

The dust opacity spectral indexes $\beta$ do not have a clear
(anti-)correlation with the disk thickness/flareness nor the disk
sizes, while larger disks are thinner and less flared.  Therefore,
it may imply that large grain settlement is not the only mechanism
for thinner and less-flared disks.  This is consistent with the
fact that protoplanetary disks are evaporated and get less flared
and thinner with time.

\citet{2009ApJ...703.1964F} argued that
protoplanetary disks with grain settlement have small $n_{13-31}$
and EW($10~\mu$m) values.
Note that EW($10~\mu$m) indicates the amount of small ($< 5~\mu$m)
silicate grains and $n_{13-31}$ represents the flareness of disks
by intervening of protostellar emission.
Unlike their argument, we did not find a strong relationship between both
properties and our flare index ($h$).
Probably, the short wavelength properties of EW($10~\mu$m) and
$n_{13-31}$ are sensitive to the warm inner and/or surface disk
region and relatively small grains, while our millimeter data are
sensitive to the cold outer and/or deep disk region and large grains.
Also, inclinations of objects may be an effective parameter, as the
light path so the optical depth along line-of-sight depends on it.
In contrast, we obtained a strong correlation between $s$ and
EW($10~\mu$m) and a strong anti-correlation between $\beta$
and $n_{13-31}$.

\subsection{Disk Substructures}

\begin{figure}
\begin{center}
\begin{tabular}{c@{\hspace{-3.5cm}}c}
\hspace{-1.0cm}
\includegraphics[angle=270,scale=0.5]{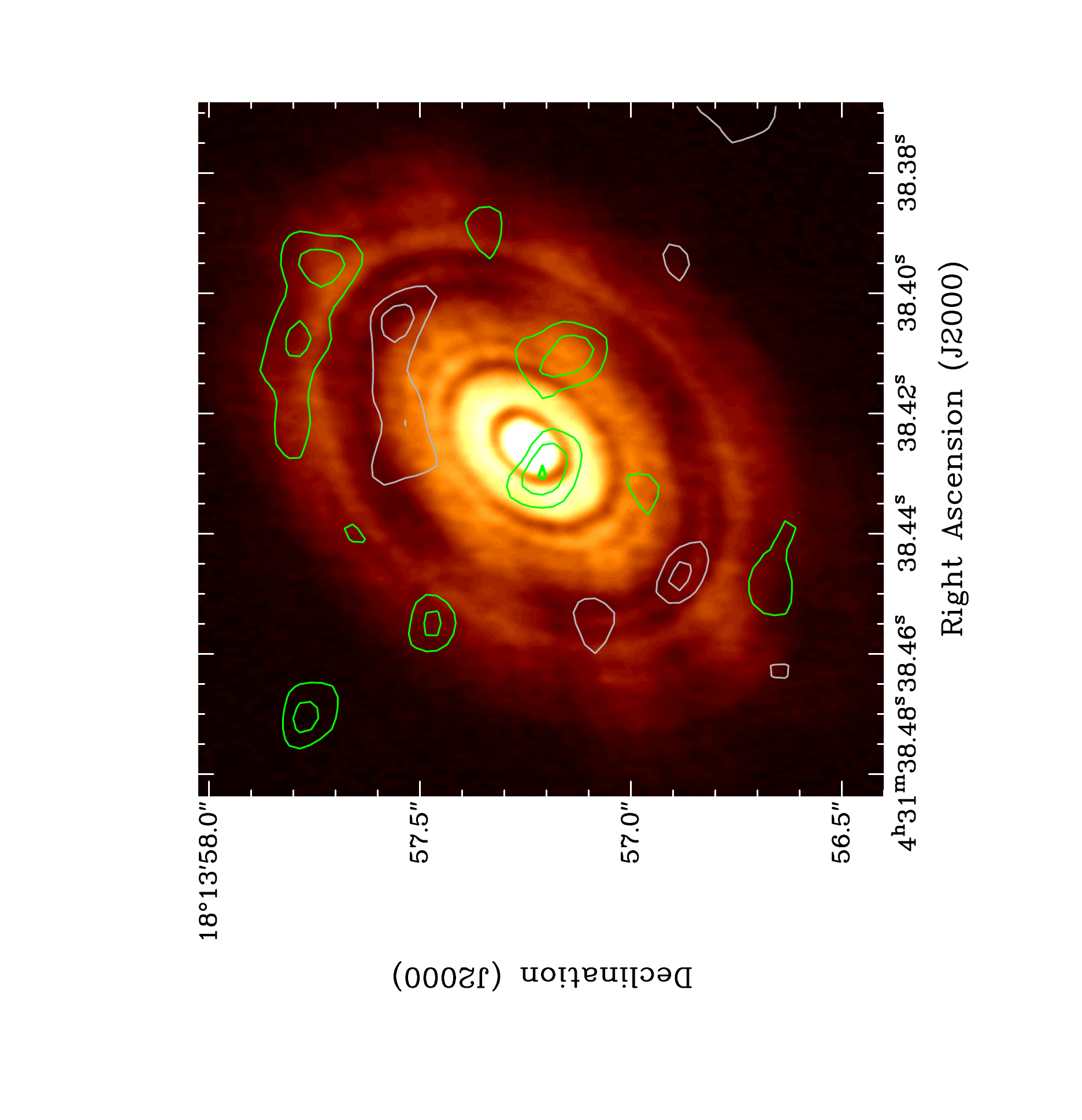} &
\includegraphics[angle=270,scale=0.5]{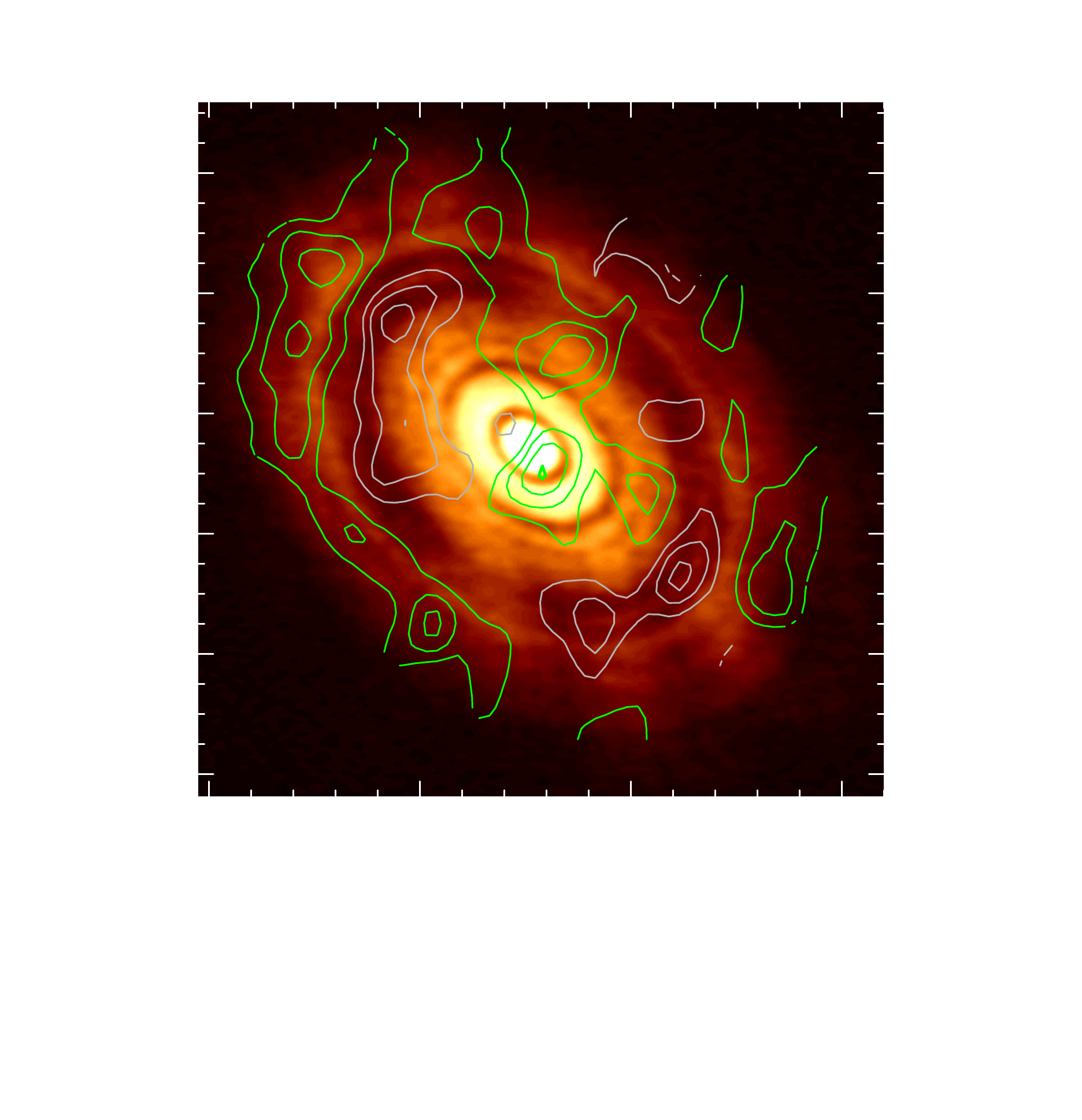}\\ 
\end{tabular}
\caption{ALMA SV data of HL Tau in Band 7 (343.5 GHz) with an angular
resolution of $0.03''\times0.02''$ ($-176\degr$) in color scales,
overlaid with the residuals of our CARMA data at \omm\ after
subtraction of the disk model in contours. The green and gray
contours indicate positive and negative levels of 2, 3, and 4 times
0.75 mJy beam$^{-1}$, respectively. The image in the right has an additional
contour of 0.75 mJy beam$^{-1}$ and the region out of the disk less than
3 mJy beam$^{-1}$ has been masked. The coordinates are in epoch
of 2014 Oct. 31, when the ALMA data were taken, and our CARMA data
were shifted by the proper motion.
%$+0.106''$ in R.A. and $-0.125''$ in Dec. 
Note that the residuals nicely match the gaps and rings.
\label{fig_hltau_alma}}
\end{center}
\end{figure}

\begin{figure}
\begin{center}
\includegraphics[angle=0,scale=0.8]{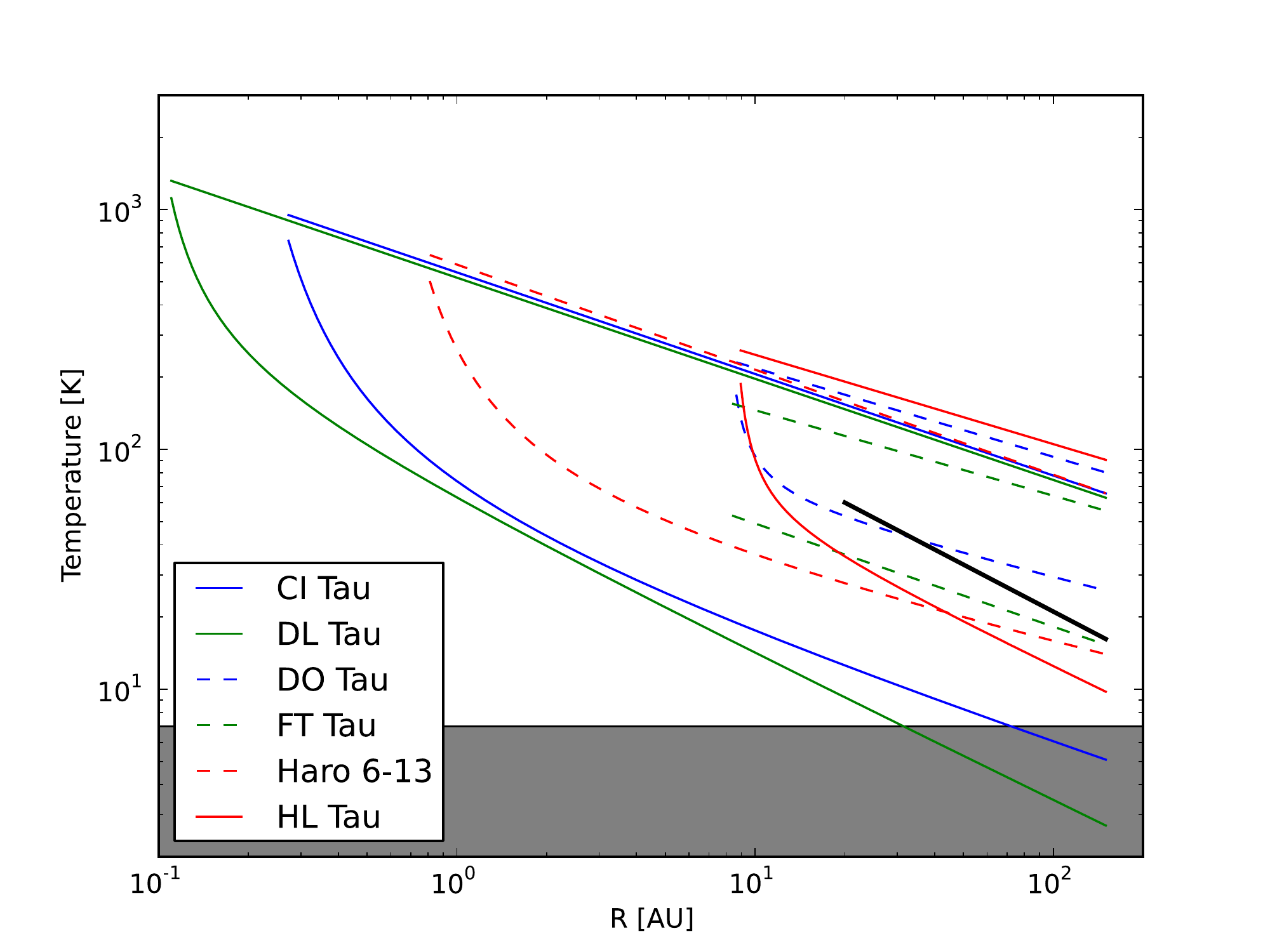}
\caption{Comparisons of temperature distributions particularly for HL Tau in
ALMA SV data and our modeling.
Radial distributions of surface ($T_s$, upper straight lines) and mid-plane
($T_m$, lower curves) temperatures are shown for each of the best fitting
accretion disk models. The black straight line indicates the power-law temperature 
distribution outlining the brightness temperature distribution of the ALMA SV data
\citep{2015arXiv150302649P}: $T_B \textrm{[K]} = 60(R/20 \textrm{AU})^{-0.65}$.
The lower limit of dust temperatures in our modeling is 7 K.
\label{fig_Tdist}}
\end{center}
\end{figure}

As discussed in \citet{2011ApJ...741....3K},
HL Tau has a region between 50 and 100 AU, which
is gravitationally unstable and could fragment.
\wktextbf{The prediction of substructures in HL Tau has been proved
by the ALMA SV data with the unprecedented angular resolution up to $0.02''$, which
resolves at least 8 gaps (and rings). Particularly, the data present a wide gap between 50 and 100 AU.
Indeed, our CARMA data show a hint of the wide ring, as shown in
Figure \ref{fig_hltau_alma}. In addition, the positive and negative
residuals nicely line up with the rings and gaps. 
%Detailed modeling for the ALMA SV data is ongoing.  
For the comparison of the CARMA residuals
with the ALMA image, we shifted CARMA data by the proper motion.
The HL Tau center positions are (R.A., Dec. in J2000) = (04:31:38.418,
+18:13:57.37) in the CARMA observations on 2009 Jan. 31 and
(04:31:38.42545, +18:13:57.242) in the ALMA observations on 2014
Oct. 30 \citep{2015arXiv150302649P}, so the proper motion is estimated
as $v_{R.A.} = 18$ mas year$^{-1}$ and $v_{decl.} = -22$ mas
year$^{-1}$ during the 5.75 year span. In addition, the temperature
distributions of our modeling are consistent with the brightness
temperature estimated by the ALMA multi-band SV data.
Figure \ref{fig_Tdist} shows examples
of our model temperature distributions in color solid and dashed
lines. In each case, the upper straight line is for the surface
temperature and the bottom curve or line is for the mid-plane temperature.
As described in Section \ref{sec_diskmodel}, the temperatures were
obtained by fitting RADMC 3D results. The black line is the outline
of the brightness temperature estimated from the ALMA multi-band
SV data \citep[Figure 3.d of ][]{2015arXiv150302649P}. 
Note that the brightness temperature is between the surface
and mid-plane temperatures of our modeling, which means that our
temperature distributions are reasonable.}

Like HL Tau, whose CARMA data show a hint and whose ALMA data provide
the detailed substructures, CI Tau and DL Tau images also show
a possible non-axisymmetric feature.
We calculated the Toomre Q parameter to check if
the three extended targets have gravitationally
unstable regions.
The Q parameter is defined as \citep{1964ApJ...139.1217T},
\begin{equation}
Q \equiv \frac{c_s \Omega}{\pi G \Sigma},
\end{equation}
where $c_s$ is the isothermal sound speed,
$\Omega$ is the orbital angular velocity ($\Omega=\sqrt{GM_*/R^3}$),
$G$ is the gravitational
constant, and $\Sigma$ is the surface density, and regions of
$Q \lesssim 1.5$ are gravitationally unstable.
As we constrained disk temperature and density distributions,
the $Q$ parameter can be calculated assuming Keplerian rotation.
Note that we utilize our mid-plane temperature distributions
for $c_s$, instead of a constant temperature: $c_s^2 = kT_m^2$.

\begin{figure}
\begin{center}
\includegraphics[angle=0,scale=0.8]{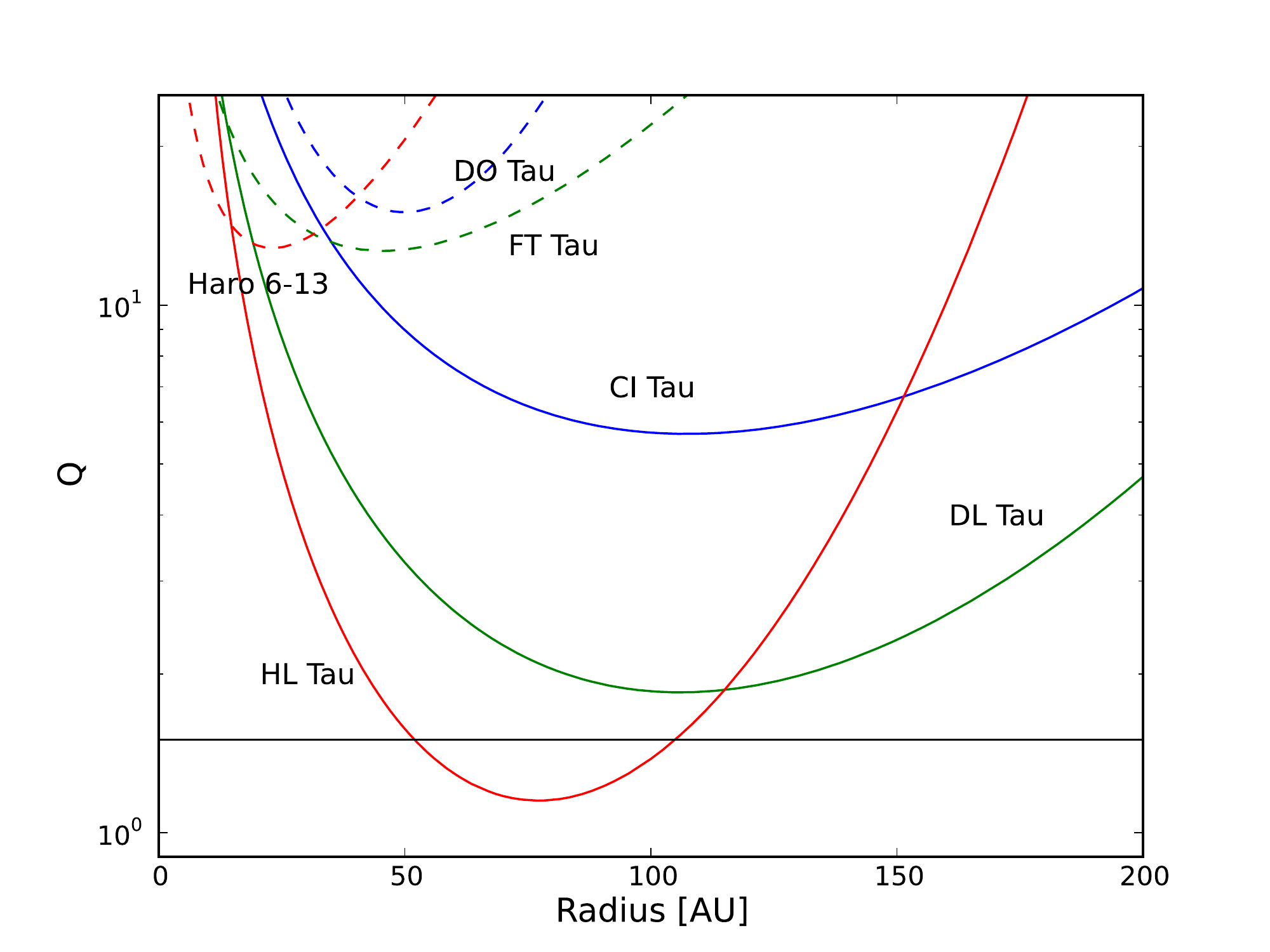}
\caption[Toomre $Q$ parameters along radius]
{The Toomre $Q$ parameter along radius in the accretion disk model.
\label{fig_qparam}}
\end{center}
\end{figure}

Figure \ref{fig_qparam} shows the $Q$ parameter along radius in the
accretion disk models.  As shown in the figure, CI Tau and DL Tau
have the smallest $Q$ minima with HL Tau, but they do not have a
region with $Q<1.5$.  However, it is possible that they could have
undetectable mass in large grains, as discussed
for the correlation between $\beta$ and disk mass in Section
\ref{sec_beta_diskmass}: in particular, $\beta$ of CI Tau is close
to 0.  On the other hand, DL Tau has a
region with the $Q$ values very close to 1.5 around 100 AU in radius,
although the sensitivity of our data does not provide any distinct
non-axisymmetric features.

Non-axisymmetric substructures are a possible place for protoplanet
formation developed by gravitational instability. Also, they allow
us to study the accretion mechanism at the late stage of star
formation.  A trailing spiral structure built by gravitational
instability is an efficient mechanism to transport angular momentum
outward \citep[e.g.,][]{2001apsf.book.....H}.  Therefore, if we
could obtain the kinematics of such a feature along, we can
study the accretion mechanism.  As the SV data proved, high angular resolution and
sensitivity observations of ALMA will clearly show whether or not
the extended disks of our sample have a substructure and
enable us to study the accretion mechanism.

While the accretion mechanism of the gravitational instability
generally appears in the outer disk region, the most successful
mechanism to explain angular momentum transport in the accretion
disks is the magnetorotational instability (MRI)
\citep[e.g.,][]{2003ARA&A..41..555B}.  The key aspects MRI requires
are magnetic fields coupled with material to act as a tension and
disk rotational velocities decreasing outward.  As with the trailing
spiral arms, in which the gravity of the mass excess acts as a
tension, a faster rotating inner region is dragged by the slower
rotating outer region connected by the magnetic field tension.
Therefore, angular momentum is transported outward.  The rotation
decreasing outward is a general property of circumstellar disks,
whose velocity appears Keplerian.  For the former aspect of magnetic
fields coupled with disk material, the key property of disks is
ionization fraction.  \citet{1996ApJ...457..355G} tackled which
parts of disks are coupled with magnetic fields and suggested a
layered accretion disk.  While the inner region ($<0.1$ AU) is
collisionally ionized, the outer region is layered with accretionally
active (ionized enough by cosmic rays) and dead zones (cold mid-plane).

The viscous accretion disk model provides a means to investigate
the physical origin of viscosity.
Viscosity of disks is normally parameterized as $\nu = \alpha c_s H$,
where $\alpha$ is a dimensionless parameter, $c_s$ is a sound speed,
and $H$ is a scale height
\citep[e.g.,][]{1973A&A....24..337S}.
On the other hand, since the mass accretion rate is related to
the viscosity in a structure-determined disk
\citep[e.g., Equation A10 in][]{2009ApJ...700.1502A},
we can estimate the $\alpha$ parameter of our disk models fitting
data, assuming an accretion rate.
For this calculation, we use the
mass accretion rate estimated by \citet{2007ApJS..169..328R} using
SED fitting (Table \ref{tab_diskacc}) and the equation derived by
\citet{2009ApJ...700.1502A}:
\begin{equation}
\alpha \approx \frac{2R_c^2}{3(2-\gamma)} \frac{\dot{M}_*}{M_d}
\Big(\frac{R}{R_c}\Big)^\gamma \frac{1}{c_s H}.
\end{equation}
Although the accretion rates determined by UV and optical spectroscopy
are more accurate \citep{2007ApJS..169..328R}, we adopted the SED
fitting results in order to have as many target values as possible.
Note that the values estimated by UV and optical spectroscopy are
available for CI Tau, DL Tau and DO Tau and are close to the minima
of the SED fitting results in the parentheses of Table \ref{tab_diskacc}.
The $\alpha$ is a function of radius but we listed only the values at 10
AU and 100 AU.  The $\alpha$ values at
10 and 100 AU of our sample disks are all in the range that MRI can
be the viscosity origin for: 0.005--0.6 \citep{2003ARA&A..41..555B}.
We assumed the accretion rate of Haro 6-13, which has not been
reported, as a typical value of $1.0\times 10^{-7}~\textrm{M}_\sun~
\textrm{year}^{-1}$.
The investigation of the viscosity origin is beyond
the scope of this study, so further attempts to understand the
exception are not made.

\section{Conclusion}

We have carried out a T Tauri disk survey using CARMA, which provides
excellent image fidelity and angular resolution.  We have
acquired multi-wavelength (\omm\ and 2.7 mm) and multi-configuration
(A, B, C, and/or D) data up to angular resolution of $0.13''$ toward
6 targets: CI Tau, DL Tau, DO Tau, FT Tau, Haro 6-13, and HL Tau.
Using visibility modeling with Bayesian inference, we obtained disk
properties of the flared viscous accretion disk model such as density
distribution, dust opacity spectral index, disk mass, disk inner
and characteristic radii, inclination, position angle, and disk
vertical structures.  In addition, we examined anti/correlations
between the properties. The power-law disk model was also employed
for investigating which disk model is preferred.

1. We found that the accretion disk model is preferred overall,
with the exception of DL Tau, which prefers the power-law disk model.
However, the understanding of the
discrepancy is not clear at the moment.
ALMA observations with a better sensitivity and angular resolution
toward a large number of protoplanetary disks are required for such
model comparison and trend investigation further.

2. We obtained correlations between the properties we constrained
using the flared accretion disk model.
Particularly, we found that disks with a steeper mid-plane density
gradient have smaller $\beta$ (large grains) and are less massive.
This suggests that grains grow and radially move and implies that
a less massive disk has a larger fraction of large grains.
In addition, we found that extended disks tend to be less flared.

3. We detected a non-axisymmetric feature toward CI Tau
with a similar pattern at both millimeter wavelengths.
However, the Toomre Q parameter criterion does not support
a gravitationally unstable structure in CI Tau.
It is interesting to note that its $\beta$ is close to 0,
which implies that the mass might be underestimated.
On the other hand, as discussed in previous studies
\citep[e.g.,][]{2011ApJ...741....3K}, HL Tau has a gravitationally
unstable region between 50 and 100 AU, where we marginally detected
a possible substructure and the ALMA SV data present a wide gap.
Also, DL Tau has a region around 100 AU with the Q values very
close to the gravitational instability regime.
ALMA observations toward these objects will provide a more clear view.

\acknowledgments
The authors thank the CARMA staff and observers for their dedicated
work and the anonymous referee for valuable comments. 
Support for CARMA construction was derived from the states
of Illinois, California, and Maryland, the James S. McDonnell
Foundation, the Gordon and Betty Moore Foundation, the Kenneth T.
and Eileen L. Norris Foundation, the University of Chicago, the
Associates of the California Institute of Technology, and the
National Science Foundation (NSF).
Ongoing CARMA development and operations
are supported by NSF under a cooperative
agreement, and by the CARMA partner universities.
L.W.L. acknowledges NSF AST-1139950.
L.G.M. acknowledges NSF AST-1139998.
This work used the Extreme Science and Engineering Discovery
Environment (XSEDE), which is supported by NSF ACI-1053575.
This paper makes use of the following ALMA data: ADS/JAO.ALMA\#2011.0.00015.SV. ALMA is a partnership of ESO (representing its member states), NSF (USA) and NINS (Japan), together with NRC (Canada), NSC and ASIAA (Taiwan), and KASI (Republic of Korea), in cooperation with the Republic of Chile. The Joint ALMA Observatory is operated by ESO, AUI/NRAO and NAOJ.

Facilities: \facility{CARMA,ALMA}

\bibliographystyle{apj}
\bibliography{kwon_disk,kwon_hltau,addedRef}

\clearpage

\clearpage
\begin{table}
\begin{small}
\caption[Disk targets]{Reference positions and stellar parameters
\label{tab_targets}}
\begin{tabular}{cccccccc}
\tableline \tableline
Targets & \multicolumn{2}{c}{Positions (J2000)\tablenotemark{a}} &
T$_{eff}$\tablenotemark{b} & L$_*$\tablenotemark{c} & M$_*$\tablenotemark{d} &
EW(10$\mu$m)\tablenotemark{e} & $n_{13-31}$\tablenotemark{e} \\
        & h m s & $\degr~'~''$ & [K] & [L$\sun$] & [M$\sun$] & \\
\tableline
CI Tau & 04 33 52.000 & +22 50 30.20 & 4060 & 1.40 & 0.70 & 2.55 & -0.17 \\
DL Tau & 04 33 39.076 & +25 20 38.14 & 4060 & 1.12 & 0.70 & 0.51 & -0.77 \\
DO Tau & 04 38 28.583 & +26 10 49.85 & 3850 & 2.70 & 0.70 & 0.94 & -0.13 \\
FT Tau & 04 23 39.178 & +24 56 14.30 & 3890\tablenotemark{1} & 0.38 & 0.70 & 1.86 & -0.46 \\
Haro 6-13 & 04 32 15.410 & +24 28 59.97 & 3850\tablenotemark{1} & 2.11\tablenotemark{1} & 0.55 & 3.87 & 0.38 \\
HL Tau & 04 31 38.471 & +18 13 58.11 & 4000\tablenotemark{2} & 8.30\tablenotemark{2} & 0.55 & -- & -- \\
\tableline
\tablenotetext{a}{Observation phase center positions. Note that the actual disk center 
positions are offset  as given in Table \ref{tab_diskdata}}.
\tablenotetext{b}{\citet{1995ApJS..101..117K}}
\tablenotetext{c}{Bolometric luminosities of \citet{1995ApJS..101..117K} are adopted.}
\tablenotetext{d}{These values are reasonably assumed, as they are not well known.
Note that they are not a sensitive parameter for modeling.}
\tablenotetext{e}{\citet{2009ApJ...703.1964F}}
\tablenotetext{1}{It is assumed as same as or similar to \citet{andrews2005}.}
\tablenotetext{2}{The values used in \citet{2011ApJ...741....3K} have been employed.}
\end{tabular}
\end{small}
\end{table}

\clearpage
\begin{table}
\begin{small}
\caption{Protoplanetary disk observations
\label{tab_diskobs}}
\begin{tabular}{llllll}
\tableline \tableline
Targets & Obs. Dates & Array & Wavelengths & Flux cal. & Gain cal.. \\
\tableline
CI Tau & 2007 Nov. 25 & B &1& MWC349 (1.9) & 0530+135 (2.4), 3C84 (4.2) \\
CI Tau & 2007 Sep. 18 & C &1& Mars         & 0530+135 (3.0) \\
CI Tau & 2008 Jun. 13 & D &1& Uranus       & 0530+135 (1.6), 3C111 (1.7) \\
CI Tau & 2008 Jan. 18 & B &3& MWC349 (1.3) & 0530+135 (3.8), 3C84 (7.1) \\
CI Tau & 2008 Oct. 29 & C &3& 3C273 (14)   & 0530+135 (2.0)  \\
CI Tau & 2009 Mar. 30 & D &3& Uranus       & 0510+180 (1.1), 3C111 (2.8) \\
\tableline
DL Tau & 2008 Dec. 11 & B &1& 3C84 (5.0)   & 0510+180 (0.9) \\
DL Tau & 2008 Dec. 14 & B &1& 3C84 (5.0)   & 0510+180 (0.9) \\
DL Tau & 2009 Mar. 11 & D &1& Uranus       & 0510+180 (0.9), 3C111 (2.3) \\
DL Tau & 2009 Jan. 21 & A &3& 3C454.3 (10) & 0510+180 (1.3), 3C111 (3.8) \\
DL Tau & 2009 Jan. 22 & A &3& 3C84 (9.0)   & 0510+180 (1.3) \\
DL Tau & 2008 Oct. 21 & C &3& Uranus       & 0530+135 (2.0), 3C111 (6.9) \\
DL Tau & 2008 Oct. 29 & C &3& Uranus       & 0530+135 (2.0), 3C111 (6.9) \\
\tableline
DO Tau & 2007 Nov. 27 & B &1& MWC349 (1.9) & 0510+180 (0.6) \\
DO Tau & 2009 Nov. 4  & C &1& Uranus       & 3C111 (1.2), 0510+180 (0.5) \\
DO Tau & 2009 Mar. 27 & D &1& Uranus       & 0510+180 (0.8), 3C111 (2.0) \\
DO Tau & 2008 Oct. 18 & C &3& 3C84 (9.0)   & 0530+135 (2.0) \\
DO Tau & 2008 Oct. 28 & C &3& Uranus       & 0530+135 (2.0), 3C111 (6.9) \\
DO Tau & 2008 Oct. 28 & C &3& 3C84 (9.0)   & 0530+135 (2.0) \\
\tableline
FT Tau & 2011 Jan. 5  & B &1& Uranus       & 3C111 (3.2), 0336+323 (0.9) \\
FT Tau & 2008 Oct. 15 & C &1& Uranus       & 0357+233 (0.35) \\
FT Tau & 2009 Feb. 15 & A &3& Uranus       & 0510+180 (1.3), 3C111 (3.5) \\
FT Tau & 2008 Oct. 26 & C &3& 3C454.3 (20) & 0530+135 (2.0), 3C111 (6.9) \\
\tableline
Haro 6-13 & 2008 Jan. 12 & B &1& 3C454.3 (11) & 0530+135 (2.2) \\
Haro 6-13 & 2008 Apr. 25 & C &1& Uranus       & 0530+135 (2.0), 3C111 (1.9) \\
Haro 6-13 & 2010 Jan. 29 & A &3& Uranus       & 3C111 (2.4), 0336+323 (2.1) \\
Haro 6-13 & 2009 Jan. 4  & B &3& Uranus       & 0510+180 (1.3) \\
Haro 6-13 & 2009 Apr. 18 & C &3& Uranus       & 0510+180 (1.0), 3C111 (2.6) \\
\tableline
HL Tau & 2009 Jan. 17 & A &1&            & 0510+180 (0.9) \\
HL Tau & 2009 Jan. 31 & A &1& 3C84       & 0510+180 (0.9) \\
HL Tau & 2009 Jan. 1  & B &1& 3C84 (5.0) & 0510+180 (0.9) \\
HL Tau & 2007 Nov. 4  & C &1& Uranus     & 0530+135 (2.6) \\
HL Tau & 2008 Feb. 16 & B &3& Uranus     & 0530+135 (4.0), 3C111 (6.7) \\
HL Tau & 2008 Oct. 27 & C &3& Uranus     & 0530+135 (2.0) \\
\tableline
\end{tabular}
\end{small}
\end{table}

\clearpage
\begin{landscape}
\begin{table}
%\begin{sidewaystable}
\begin{small}
\caption{Reduced data sets. The total fluxes are estimated as a flux
sum over the target region beyond 2$\sigma$ levels and their uncertainties
are statistical, which do not include the absolute flux calibration
uncertainties.  The $\Delta$RA and $\Delta$Dec are offsets of target positions
from the phase centers.
\label{tab_diskdata}}
\begin{tabular}{lccr@{---}llcr@{ $\pm$ }llcc}
\tableline \tableline
\multicolumn{1}{c}{Targets} & Arrays & \multicolumn{1}{c}{Freq.} &
\multicolumn{2}{c}{{\it uv} coverage} & \multicolumn{1}{c}{Beam} &
RMS & \multicolumn{2}{c}{Total flux} & \multicolumn{1}{c}{Gaussian fit sizes} &
$\Delta$RA & $\Delta$Dec \\
      &      & [GHz]     & \multicolumn{2}{c}{[k$\lambda$]}  &
\multicolumn{1}{c}{[$''\times''$ (PA\degr)]} & [mJy beam$^{-1}$] &
\multicolumn{2}{c}{[mJy]} &
\multicolumn{1}{c}{[$''\times''$ (PA\degr)]} & [$''$]  & [$''$] \\
\tableline
CI Tau & BCD & 229 & 6.8&726.5 & $0.45\times0.31$ (-81) & 1.8 & 97.8&5.7 & $0.82\times0.64$ (-14) & 0.24 & -0.20 \\
       & BCD & 113 & 3.9&363.8 & $0.95\times0.69$ (89) & 0.55 & 28.3&1.5 & $1.04\times0.92$ (13) \\
DL Tau & BD & 229 & 8.5&723.8 & $0.49\times0.29$ (82) & 0.97 & 194.2&4.8 & $1.02\times0.78$ (59) & 0.05 & -0.26 \\
       & AC & 113 & 7.4&727.6 & $0.31\times0.27$ (-83)& 0.57 & 35.0&2.4  & $1.14\times1.11$ (9) \\
DO Tau & BCD & 229 & 7.0&616.6 & $0.48\times0.42$ (80) & 1.2 & 125.4&3.5 & $0.59\times0.47$ (89) & 0.07 & -0.58 \\
       & C  & 113 & 8.3&143.8 & $1.75\times1.19$ (83) & 0.67 & 30.8&1.3  & point source \\
FT Tau & BC & 229 & 17.0&620.0 & $0.44\times0.29$ (-71) & 1.1 & 104.1&3.3 & $0.46\times0.37$ (-52) & 0.20 & -0.30 \\
       & AC & 113 & 4.1&727.6 & $0.31\times0.28$ (-9) & 0.53 & 28.8&2.0 & $1.27\times0.85$ (72) \\
Haro 6-13 & BC & 229 & 12.3&616.6 & $0.54\times0.39$ (-54) & 2.3 & 111.2&4.9 & $0.34\times0.28$ (-7)&0.15 & -0.59 \\
          & ABC  & 113 & 8.0&727.6 & $0.34\times0.29$ (-84) & 0.48 & 28.7&1.0 & $0.26\times0.20$ (-69) \\
HL Tau & ABC & 229 & 15.5&1452.0 & $0.17\times0.13$ (85) & 0.75 & 685.0&6.7 & $0.85\times0.64$ (-46)&-0.75&-0.74 \\
       & BC  & 112 & 8.0&359.0 & $1.01\times0.67$ (73) & 0.93 & 118.4&2.6 & $0.84\times0.71$ (-23) \\
\tableline
\end{tabular}
\end{small}
%\end{sidewaystable}
\end{table}
\end{landscape}

\clearpage
\begin{landscape}
\begin{table}
\begin{footnotesize}
\caption[Disk fitting results]{Disk fitting results
of free parameters.
\label{tab_parampost}}
\begin{tabular}{c|ccccccc|cc}
\tableline \tableline
\multicolumn{10}{c}{Viscous accretion disk model}\\
\tableline
Targets & $s$ & $\beta$ & $M_{disk}$ & $R_{in}$ & $R_{c}$ & $\theta_i$ & PA &
$\gamma$\tablenotemark{a} & $R_t$ \\
       &    &    & [\msun] & [AU] & [AU] & [$\degr$] & [$\degr$] &  & [AU]\\
\tableline
CI Tau  & 1.67$\pm$0.24  & 0.049$\pm$0.027  & 0.018$\pm$0.002  & 4.2$\pm$2.6  & 119.6$\pm$6.1  & 61.0$\pm$2.9  & 17.6$\pm$2.6 &  (0.72) & (57.4) \\
DL Tau  & 1.22$\pm$0.14  & 0.408$\pm$0.022  & 0.041$\pm$0.005  & 1.3$\pm$1.1  & 112.2$\pm$2.4  & -38.6$\pm$1.3  & 55.2$\pm$2.0 &  (0.27) & (54.8) \\
DO Tau  & 1.38$\pm$0.21  & 0.029$\pm$0.018  & 0.014$\pm$0.001  & 6.7$\pm$1.6  & 54.2$\pm$1.1  & -32.5$\pm$2.0  & 89.7$\pm$3.8 &  (-0.17) & (27.6) \\
FT Tau  & 1.85$\pm$0.24  & 0.384$\pm$0.024  & 0.021$\pm$0.002  & 8.09$\pm$0.98  & 55.2$\pm$2.0  & 33.8$\pm$1.7  & 136.1$\pm$2.9 &  (0.60) & (26.5)\\
Haro 6-13  & 2.16$\pm$0.21  & -0.152$\pm$0.024  & 0.012$\pm$0.001  & 1.6$\pm$1.3  & 28.9$\pm$2.1  & 41.5$\pm$3.3  & 167.2$\pm$5.8 &  (1.06) & (14.8)\\
HL Tau  & 0.7483$\pm$0.0087  & 0.6745$\pm$0.0069  & 0.105$\pm$0.001  & 8.78$\pm$0.19  & 80.20$\pm$0.34  & 41.19$\pm$0.22  & 135.38$\pm$0.34 &  (-0.20) & (40.9)\\
\tableline
\multicolumn{10}{c}{Power-law disk model}\\
\tableline
Targets & $s$ & $\beta$ & $M_{disk}$ & $R_{in}$ & $R_{out}$ & $\theta_i$ & PA &
$p$\tablenotemark{b}  \\
       &    &    & [\msun] & [AU] & [AU] & [$\degr$] & [$\degr$] &  \\
\tableline
CI Tau  & 2.15$\pm$0.25  & 0.036$\pm$0.027  & 0.017$\pm$0.002  & 6.5$\pm$2.0  & 195$\pm$11  & 62.5$\pm$2.6  & 17.7$\pm$2.5 &  (1.05) \\
DL Tau  & 2.08$\pm$0.23  & 0.387$\pm$0.021  & 0.029$\pm$0.003  & 8.18$\pm$0.89  & 152.6$\pm$2.3  & -38.4$\pm$1.3  & 55.6$\pm$2.0 & (0.53)  \\
DO Tau  & 1.98$\pm$0.23  & 0.021$\pm$0.018  & 0.013$\pm$0.001  & 7.3$\pm$1.5  & 75.4$\pm$2.0  & -32.8$\pm$2.0  & 90.7$\pm$3.5 &  (0.43) \\
FT Tau  & 2.80$\pm$0.15  & 0.341$\pm$0.024  & 0.015$\pm$0.002  & 8.70$\pm$0.27  & 98.2$\pm$5.1  & 33.2$\pm$1.7  & 136.0$\pm$3.0 &  (1.25)\\
Haro 6-13  & 2.76$\pm$0.21  & -0.164$\pm$0.024  & 0.011$\pm$0.002  & 1.54$\pm$0.65  & 61.9$\pm$5.2  & 43.5$\pm$2.9  & 167.3$\pm$5.9 &  (1.81)\\
HL Tau  & 2.14$\pm$0.01  & 0.615$\pm$0.006  & 0.067$\pm$0.001  & 8.70$\pm$0.02  & 109.71$\pm$0.23  & 40.9$\pm$0.20  & 135.7$\pm$0.34 & (0.59) \\
\tableline
\end{tabular}
\tablecomments{The presented uncertainties are statistical fit errors.
As discussed in Table 1 of \citet{2011ApJ...741....3K},
the systematic uncertainties are about 0.07 in $s$ due to the flux
calibration errors over different configuration data,
0.25 in $\beta$ due to the flux calibration uncertainties over wavelengths,
and 10\% in $M_{disk}$ due to the flux calibration uncertainty at a
dominant wavelength. The systematic uncertainties of $R_{in}$, $R_c$
and $R_{out}$ depend on the angular resolution.  The $R_{in}$ uncertainty
is also affected by the data sensitivity.
}
\tablenotetext{a}{For computing this ($s-h$), the best h values of the
accretion disk model are adopted: 0.95 (CI Tau, DL Tau, HL Tau), 1.55 (DO
Tau), 1.25 (FT Tau), and 1.10 (Haro 6-13). \textbf{It is not intended to
provide the best fitting result for surface density distributions in this paper. These are only
for a rough comparison with other studies.}}
\tablenotetext{b}{The best h values of the power-law disk model are adopted: 1.10
(CI Tau), 1.55 (DL Tau, DO Tau, FT Tau, HL Tau), and 0.95 (Haro 6-13).}
\end{footnotesize}
\end{table}
\end{landscape}

\clearpage
\begin{table}
\begin{small}
\caption[Disk model comparison]{Disk model comparison.
\label{tab_modelcomp}}
\begin{tabular}{ccccl}
\tableline \tableline
Targets & ln($P(D\mid H_A)$) & ln($P(D\mid H_P)$) & ln($K$) & Preferable model\\
\tableline
CI Tau     & -1221129.9 & -1221128.1 & 0.8 & Comparable \\
DL Tau     & -5336360.9 & -5336365.6 & -4.7 & Power-law \\
DO Tau     & -1794530.0 & -1794521.3 & 8.7 & Accretion \\
FT Tau     & -5444060.7 & -5444044.3 & 16.4& Accretion \\
Haro 6-13  & -3053420.6 & -3053420.1 & 0.5 & Comparable \\
HL Tau     & -7863636.4 & -7863346.8 & 289.6 & Accretion \\
\tableline
\end{tabular}
\end{small}
\end{table}

\clearpage
\begin{table}
\begin{small}
\caption[Correlation coefficients between YSO properties]{
Correlation coefficients between properties.
HL Tau is not included in the relationships with EW($10~\mu$m) and $n_{13-31}$
and the bold-font anti/correlations are plotted in Figure \ref{fig_correl}.
\label{tab_correl}}
\begin{tabular}{crrrrr|rr}
\tableline \tableline
Properties & \multicolumn{1}{c}{$\beta$} & $M_{disk}$ & $R_{in}$ & $R_{c}$ &
\multicolumn{1}{c}{$h$} & EW & $n_{13-31}$ \\
\tableline
$s$        & \bf{-0.77}& \bf{-0.84}& -0.41& -0.47&       0.21&  0.93&  0.74 \\
$\beta$    &      1.00 &  \bf{0.84}&  0.52&  0.36&      -0.36& -0.70& -0.95 \\
$M_{disk}$ &           &  1.00     &  0.44&  0.26&      -0.48& -0.65& -0.86 \\
$R_{in}$   &           &           &  1.00& -0.16&       0.34& -0.23& -0.13 \\
$R_{c}$    &           &           &      &  1.00& \bf{-0.59}& -0.44& -0.63 \\
$h$        &           &           &      &      &       1.00& -0.25&  0.19 \\
\tableline
EW($10~\mu$m)&         &           &      &      &           &  1.00&  0.83 \\
$n_{13-31}$&           &           &      &      &           &      &  1.00 \\
\tableline
\end{tabular}
\end{small}
\end{table}

\clearpage
\begin{table}
\begin{small}
\caption[Disk mass accretion properties]{Disk mass accretion properties.
\label{tab_diskacc}}
\begin{tabular}{ccccc}
\tableline \tableline
Targets & $\dot{M}_{disk}$\tablenotemark{a} &
$M_{disk}\tablenotemark{b}/\dot{M}_{disk}$ &
$\alpha$(10 AU) & $\alpha$(100 AU) \\
 & [$\times 10^{-7}$ \msun\ year$^{-1}$] & [$\times 10^5$ year] \\
\tableline
CI Tau & 1.2 (0.49--2.6) & 1.5 & 0.43  & 0.18 \\
DL Tau  & 1.2 (0.69--1.2) & 3.5 & 0.30  & 0.098 \\
DO Tau  & 6.8 (1.5--6.8) & 0.2 & 0.37  & 0.017 \\
FT Tau  & 0.51 (0.028--1.6) & 4.0 & 0.029 & 0.014 \\
Haro 6-13 & 1.0\tablenotemark{1} & 1.2  & 0.018 & 0.013 \\
HL Tau  & 10 (3.0--51) & 1.0 & 0.41 & 0.079 \\
\tableline
\end{tabular}
\tablenotetext{a}{These values come from \citet{2007ApJS..169..328R}.
Particularly, we adopt the SED fitting values, which are available for
most of our targets: the best fitting value and the ranges in parentheses.
}
\tablenotetext{b}{The accretion disk model estimates in Table
\ref{tab_parampost} are used.}
\tablenotetext{1}{Assumed as it is not available.}
\end{small}
\end{table}

\end{document}